\documentclass[aps,prc,a4paper,nofootinbib,showpacs,showkeys,
preprintnumbers,superscriptaddress,twocolumn]{revtex4}

\usepackage{amsmath}
\usepackage{amssymb}
\usepackage{bm}
\usepackage{graphicx}
\usepackage{color}

\begin{document}

\title{Charmed Hadrons from Coalescence plus Fragmentation \\ in relativistic nucleus-nucleus collisions at RHIC and LHC}

\author{S. Plumari}
\affiliation{Department of Physics and Astronomy, University of Catania, Via S. Sofia 64, 1-95125 Catania, Italy}

\author{V. Minissale}
\affiliation{Department of Physics and Astronomy, University of Catania, Via S. Sofia 64, 1-95125 Catania, Italy}
\affiliation{Laboratori Nazionali del Sud, INFN-LNS, Via S. Sofia 62, I-95123 Catania, Italy}

\author{S.K. Das}
\affiliation{School of Nuclear Science and Technology, Lanzhou University, 730000 Lanzhou, China}
\affiliation{Department of Physics and Astronomy, University of Catania, Via S. Sofia 64, 1-95125 Catania, Italy}

\author{G. Coci}
\affiliation{Department of Physics and Astronomy, University of Catania, Via S. Sofia 64, 1-95125 Catania, Italy}
\affiliation{Laboratori Nazionali del Sud, INFN-LNS, Via S. Sofia 62, I-95123 Catania, Italy}

\author{V. Greco}
\affiliation{Department of Physics and Astronomy, University of Catania, Via S. Sofia 64, 1-95125 Catania, Italy}
\affiliation{Laboratori Nazionali del Sud, INFN-LNS, Via S. Sofia 62, I-95123 Catania, Italy}

\begin{abstract}
In a coalescence plus fragmentation approach we calculate the heavy baryon/meson 
ratio and the $p_T$ spectra of charmed hadrons $D^{0}$, $D_{s}$ and  $\Lambda_{c}^{+}$ 
in a wide range of transverse momentum from low $p_T$ up to 
about 10 GeV and discuss their ratios from RHIC to 
LHC energies without any change of the coalescence parameters.
We have included the contribution from decays of heavy hadron resonances and also 
the one due to fragmentation of heavy quarks which do not undergo the coalescence process. 
The coalescence process is tuned to have all charm quarks 
hadronizing in the $p_T\rightarrow 0$ limit and at finite $p_T$  charm quarks not 
undergoing coalescence are hadronized by independent fragmentation.
The $p_T$ dependence of the  baryon/meson ratios are found 
to be sensitive to the masses of coalescing quarks, in particular the 
$\Lambda_{c}/D^{0}$ can reach values of about $\rm 1\div 1.5 $ at $p_T \approx \, 3$ \mbox{GeV}, 
or larger, similarly to the light baryon/meson ratio 
like $p/\pi$ and $\Lambda/K$, however a marked difference is a quite weak $p_T$ dependence 
with respect to the light case, such that a larger value at intermediate $p_T$ implies a 
relatively large value also for the integrated yields. A comparison with other 
coalescence model and with the prediction 
of thermal model is discussed.
\end{abstract}



\maketitle

\section{Introduction}
Ultra-relativistic heavy ion collisions at Large Hadron Collider 
(LHC) and at Relativistic Heavy-Ion Collider (RHIC) have been 
designed to reach a new state of matter composed of a strongly 
interacting plasma of deconfined quark and gluons, the so called 
Quark-Gluon Plasma (QGP). Such a form of matter should have permeated the early universe
in the first microseconds during its expansion \cite{Castorina:2015ava}, and it is 
still an open compelling question what was the role of unknown new physics interactions on the
stability condition that allowed the universe itself to evolve into such a state 
\cite{Branchina:2014usa,Branchina:2015nda,Branchina:2016bws}.

The matter created on Earth at RHIC and LHC have revealed many 
interesting and surprising phenomena such as strong collective 
flows of the final state particle suggesting that the system 
created behaves like an almost perfect fluid with a very small 
shear viscosity to entropy density ratio as suggested by 
different theoretical calculations
\cite{Romatschke:2007mq,Schenke:2011bn,Gale:2012rq,Ruggieri:2013bda,Ruggieri:2013ova,Plumari:2015cfa}. 
The bulk properties of the matter created are governed by the light quarks and gluons 
while heavy quarks like charm or bottom quarks are useful probes 
of the QGP properties 
\cite{Linnyk:2008hp,He:2012df,He:2011qa,Uphoff:2012gb,Cao:2015hia,Nahrgang:2014vza,Scardina:2017ipo,Das:2017dsh,Das:2016cwd,Das:2015ana,Das:2013kea,Tolos:2016slr,Cao:2016gvr,Beraudo:2017gxw,Beraudo:2014boa,Alberico:2011zy}.
In their final state the charm quarks appear as constituent of charmed 
hadrons mainly $D$ mesons and $\Lambda_{c}$, $\Sigma_{c}$ baryons. 

The experimental advances in the study  not only of heavy mesons like $D$ mesons 
but also of $\Lambda_{c}$ baryons are important in order to have a new insight in understanding the 
hadronization mechanism in the QGP. Indeed at the energy of LHC and RHIC even in p+p collisions
the available experimental data are poor and a clear understanding of the hadronization mechanism is missing.
In AA collisions a systematic study of the baryon 
over meson ratio for different species from light to heavy flavor can 
permit to shed light on the underlying microscopic hadronization mechanism. 
For light and strange hadrons there is an enhancement of the 
baryon over meson ratio compared to the one for p+p collision is seen \cite{Greco:2003xt,Fries:2003vb,Greco:2003mm,Fries:2008hs,Minissale:2015zwa}. In 
particular it has been found that this ratio in nucleus-nucleus collision has 
a shape with a peak around $p_{T}\simeq 3 \, \mbox{GeV}$ with 
$p/\pi^{+}$, $\overline{p}/\pi^{-}$ and $\Lambda/K^{0}$ about 1 which is a 
factor 3 larger with respect to the one in p+p collisions.
Recent experimental results from STAR collaboration have shown that a
similar value of the baryon/meson ratio is expected in the heavy 
flavor sector \cite{Dong:2017dws,Xie:2017jcq,Zhou:2017ikn}. In particular the experimental data in 
$10-60 \%$ central $Au+Au$ collisions have shown a $\Lambda_c/D^0 \simeq 1.3 \, \pm \, 0.5$ for 
$3 < p_{T} < 6 \, \mbox{GeV}$ which is a very large enhancement compared to the value predicted by the
charm hadron fragmentation ratio or by the PYTHIA for p+p collisions \cite{Lisovyi:2015uqa,Sjostrand:2006za}.
Also such a ratio is quite larger than the predictions for the integrated yield within the the statistical hadronization model (SHM)
where $\Lambda_{c}/D^{0} \, \simeq 0.25 - 0.3$ \cite{Kuznetsova:2006bh,Andronic:2007zu,Andronic:2010dt}.

The idea of the coalescence model comes from the fact that comoving partons in the 
quark-gluon plasma combine their transverse momentum to produce a final-state
meson or baryon with higher transverse momentum and it was initially suggested in Ref.s 
\cite{Greco:2003xt,Fries:2003vb,Greco:2003mm,Fries:2003kq,Molnar:2003ff}.
In these first papers on quark coalescence, the mechanism is applied to explain 
succesfully the different $p_T$ spectra and the splitting of elliptic flow 
in a meson and a baryon branch. 
Afterwards, the quark coalescence model has been extended to include finite width that take into account 
for off-shell effects which allow to include the constraint of energy conservation 
\cite{Ravagli:2007xx,Ravagli:2008rt,Cassing:2009vt}. More recently it has been extended to LHC energies, 
including more resonances and correctly describing the spectra of main light hadrons like $\pi, K, p, \phi, \Lambda$
and in particular the baryon-to-meson ratios \cite{Minissale:2015zwa}.

In the heavy quark sector, in particular for charm flavour there has been a even more general consensus 
on the key role of an hadronization by coalescence to correctly predict the 
$p_T$ spectra and the $v_2$ of D mesons \cite{Greco:2017rro,vanHees:2005wb,Song:2015ykw,Cao:2013ita,Greco:2003vf}.
Instead only few studies have investigated the modification 
of the relative abundance of the different heavy hadron species produced. 
In particular this can manifest in a baryon-to-meson enhancement for charmed hadrons.
Large value for $\Lambda_{C}/D^0$ due to coalescence was first suggested in \cite{Lee:2007wr,Oh:2009zj}
where, based on di-quark or three-quark coalescence mechanism with full thermalized 
charm quarks, the predicted $\Lambda_{c}/D^{0}$ ratio is found to be comparable to 
the recent experimental data by STAR.
Other predictions with lower values at intermediate $p_T$, $\Lambda_{C}/D^0 \approx \,0.4$, were presented 
by some of the authors in Ref.s \cite{Ghosh:2014oia,Das:2016llg}.
In the following, and in particular in section \ref{section:RHIC}, we aim also at clarifying the reasons behind
such different predictions.

In this paper, we employ the covariant coalescence approach developed in 
\cite{Greco:2003mm,Minissale:2015zwa} which is based on the phase-space quark 
coalescence as done in \cite{Fries:2003vb,Fries:2003kq}. We solve the multidimensional 
integral in the coalescence formula by a Monte Carlo approach including a 3D geometry 
of the fireball. In this approach a radial flow correlation in the partonic spectra 
is included and a charm distribution function in $p_T$ from realistic simulation of heavy-ion collision 
have been considered \cite{Scardina:2017ipo}. 

We calculate the transverse momentum spectra of $D$ 
mesons and $\Lambda_{c}$ and the $p_{T}$ dependence of the baryon over meson 
ratio with coalescence and fragmentation.
In addition to the direct formation of D mesons and $\Lambda_{c}$ we have 
also included the main contribution from first excited states.
In particular we have considered the decay of $D^*$ mesons for $D^0$ production 
and $\Sigma_{c}$ and $\Sigma_{c}^*$ baryon decays for $\Lambda_{c}$ production.

This paper is organized as follows.  
In section \ref{section:Coal}, we introduce the general formalism of the covariant 
coalescence model used in this work for both light and heavy
quarks. In particular, we discuss how the width parameters in the
Wigner function have been fixed and the numerical method used to solve 
the coalescence integrals.
In section \ref{section:Fragm}, we discuss the hadronization by fragmentation and how 
in our model we include both hadronization by fragmentation and coalescence.
In section \ref{section:Thermal} we use a simple thermal model to show the role of
heavy hadron resonances in the $\Lambda_C/D^0$ ratio.
How fireball size, mini-jets and the quark-gluon plasma partons are determined 
is described in section \ref{section:Fireball}. 
In section \ref{section:RHIC} 
results for the transverse momentum spectra of $D$ mesons and $\Lambda_C$ baryons 
obtained from the coalescence model and the $p_T$ dependence of the 
baryon/meson ratio have been described  for RHIC energy. Also a direct comparison between the
present approach and the one in \cite{Oh:2009zj} is presented.
In section \ref{section:LHC} for LHC energy.
Finally, we conclude in section \ref{section:Conclusion} with a summary of the present work.

\section{Coalescence model}
\label{section:Coal}
We start this section by recalling the basic elements of the coalescence model developed in \cite{Greco:2003mm,Greco:2003vf,Fries:2003kq,Fries:2003vb}
and based on the Wigner formalism that in its original 
version was developed for nucleon coalescence \cite{Dover:1991zn}.
The momentum spectrum of hadrons formed by coalescence of quarks can be written as:
\begin{eqnarray}
\label{eq-coal}
\frac{dN_{H}}{dyd^{2}P_{T}}&=& g_{H} \int \prod^{n}_{i=1} \frac{d^{3}p_{i}}{(2\pi)^{3}E_{i}} p_{i} \cdot d\sigma_{i}  \; f_{q_i}(x_{i}, p_{i})\nonumber \\ 
&\times& f_{H}(x_{1}...x_{n}, p_{1}...p_{n})\, \delta^{(2)} \left(P_{T}-\sum^{n}_{i=1} p_{T,i} \right)
\end{eqnarray}
where $d\sigma_{i}$ denotes an element of a space-like hypersurface, 
$g_{H}$ is the statistical factor to form a colorless hadron from quarks and antiquarks with spin 1/2 
while $f_{q_i}$ are the quark (anti-quark) phase-space distribution functions for i-th quark (anti-quark). 
Finally $f_{H}(x_{1}...x_{n}, p_{1}...p_{n})$ is the Wigner function and 
describes the spatial and momentum distribution of quarks in a hadron
and can be, generally, directly related to the hadron wave function.
For $n=2$ Eq. \ref{eq-coal} describes meson formation, while for $n=3$ the baryon one.
For $D$ mesons considering the spin, color and flavor 
statistical factors giving the probability that \textit{n} random quarks have the right colour, spin, isospin 
matching the quantum number of the considered hadron, for D meson the factor is $g_{D}=1/36$. 
For baryons considered in present study, i.e. $\Lambda_c$ the statistical factors is $g_{\Lambda}=1/108$.

Following the Ref. \cite{Greco:2003vf,Oh:2009zj} we adopt for the Wigner distribution function a Gaussian shape in space and momentum, 
\begin{eqnarray} 
f_{M}(x_{1}, x_{2}; p_{1}, p_{2})&=&A_{W}\exp{\Big(-\frac{x_{r1}^2}{\sigma_r^2} - p_{r1}^2 \sigma_r^2\Big)}
\label{Eq:Wigner_M}
\end{eqnarray}
where the 4-vectors for the relative coordinates in space and momentum $x_{r1}$ and $p_{r1}$ are related to the quark coordinate 
in space and momentum by the Jacobian transformations:
\begin{eqnarray} 
x_{r1}&=&x_{1} - x_{2} \nonumber \\
p_{r1}&=&\frac{m_{2} p_{1}- m_{1} p_{2}}{m_{1}+m_{2}}
\label{Eq:JACOBI1}
\end{eqnarray}
In Eq. (\ref{Eq:Wigner_M}) and  (\ref{Eq:Wigner_B}) $A_{W}$ is a normalization constant fixed to guarantee that in the limit $p \to 0$
we have all the charm hadronizing.
While $\sigma_r$ is the covariant width parameter, it can be related
to the oscillator frequency $\omega$ by $\sigma=1/\sqrt{\mu \omega}$ where $\mu=(m_1 m_2)/(m_1+m_2)$ is the reduced mass.
The width of $f_M$ is linked to the size of the hadron and in particular to the root mean 
square charge radius of the meson by
\begin{eqnarray} 
\langle r^2\rangle_{ch}&=&Q_1\langle(x_1-X_{cm})^2\rangle+Q_2\langle(x_2-X_{cm})^2\rangle \nonumber \\
&=&\frac{3}{2}  \frac{Q_1 m_2^2+Q_2 m_1^2}{(m_1+m_2)^2} \sigma_r^{2}
\end{eqnarray}
with $Q_i$ the charge of the i-th quark and the center-of-mass 
coordinate calculated as \\
$X_{cm}=\sum_{i=1}^{2} m_i x_i/\sum_{i=1}^2 m_i$ .\\

For a baryon we have a similar Wigner function expressed in term of the appropriate relative coordinates:
\begin{eqnarray}
\label{Eq:Wigner_B}
f_B&=& A_{W}\exp{\Big(-\frac{x_{r1}^2}{\sigma_{r1}^2} - p_{r1}^2 \sigma_{r1}^2\Big)} \nonumber\\
&\times& A_{W} \exp{\Big(-\frac{x_{r2}^2}{\sigma_{r2}^2} - p_{r2}^2 \sigma_{r2}^2\Big)} 
\end{eqnarray} 
where the 4-vectors for the relative coordinates $x_{r1}$ and $p_{r1}$ are the same as in Eq. (\ref{Eq:JACOBI1})
while $x_{r2}$ and $p_{r2}$ are given by the Jacobian trasformations::
\begin{eqnarray} 
x_{r2}&=&\frac{m_1 x_1 +m_2 x_2}{m_1+m_2}-x_3 \nonumber \\
p_{r2}&=&\frac{m_{3} (p_{1}+p_{2})- (m_{1} +m_{2})p_{3}}{m_{1}+m_{2}+m_{3}}
\end{eqnarray}
With the normalization factor given by $A_{W}$. The width parameters $\sigma_{ri}$ are given by
$\sigma_{ri}=1/\sqrt{\mu_{i} \omega}$ where $\mu_{i}$ are the reduced masses
\begin{eqnarray} 
\mu_1=\frac{m_1 m_2}{m_1+m_2}, & & \mu_2= \frac{(m_1+ m_2)m_3}{m_1+m_2+m_3}
\end{eqnarray}
In a similar way to the mesons, the oscillator frequency can be related to the root mean square charge radius of the baryons by
\begin{eqnarray} 
\langle r^2\rangle_{ch}&=&\sum_{i=1}^{3}Q_i\langle(x_i-X_{cm})^2\rangle = \frac{3}{2} \frac{m_2^2 Q_1+m_1^2 Q_2}{(m_1+m_2)^2} \sigma_{r 1}^2  \nonumber \\ 
&+&\frac{3}{2} \frac{m_3^2 (Q_1+Q_2)+(m_1+m_2)^2 Q_3}{(m_1+m_2+m_3)^2} \sigma_{r 2}^2 
\end{eqnarray}

The width parameters $\sigma_{r \, 1,2}$ in the Wigner functions for mesons and baryons should depend on the hadron species
and can be calculated from the charge radius of the hadrons according to quark model \cite{Hwang:2001th,Albertus:2003sx}.
This will be our main choice even if we will discuss what is the effect  of other choices like in \cite{Ghosh:2014oia} and in \cite{Oh:2009zj}.
In particular in \cite{Minissale:2015zwa} by some of the present author a different choice was made and will be discussed together with all the results shown in Fig. \ref{Fig:LambdaD0_ratio}.

We note that the Wigner function for the D mesons has only one parameter $\sigma_r$ that we fix in order 
to have the mean square charge radius of $D^{+}$ meson $\langle r^2\rangle_{ch}=0.184 \, \mbox{fm}^{2}$ 
corresponding to a $\sigma_p=\sigma_r^{-1}=0.283 \, \mbox{GeV}$.
For $\Lambda_{c}^{+}$ the widths are fixed by the mean square charge radius of 
$\Lambda_{c}^{+}$ which is given by $\langle r^2\rangle_{ch}=0.15 \, \mbox{fm}^{2}$.
Notice that also for baryons there is only one free parameter, because 
the two widths are related by the oscillatory frequency $\omega$ 
through the reduced masses $\sigma_{p i}=\sigma_{r i}^{-1}=1/\sqrt{\mu_{i} \omega}$.
The corresponding widths are $\sigma_{p 1}= 0.18 \, \mbox{\mbox{GeV}}$ and $\sigma_{p 2}=0.342 \, \mbox{GeV}$. 

Numerically, the multi-dimensional integrals in the coalescence formula are evaluated by the 
Monte-Carlo method shown in \cite{Greco:2003mm}. We introduce a large 
number of test partons with uniform distribution in the transverse plane and rapidity $y_{z}$, then in momentum space we associate a probability $P_q(i)$ to the i-th test parton with momentum ${\bf p}_{\rm T}(i)$, proportionally to the parton momentum distribution at ${\bf p}_{\rm T}$.
The proportionality is given by a constant that normalize the sum of all parton probabilities to the total parton number.

Once normalized, using test partons, the coalescence formulas for mesons can be re-written as
\begin{eqnarray}
\frac{dN_M}{d^2{\bf p}_{\rm T}}&=&g_M
\sum_{i,j}P_q(i)P_{\bar q}(j)\delta^{(2)}({\bf p}_{\rm T}-{\bf p}_{i{\rm T}}
-{\bf p}_{j{\rm T}})\nonumber\\
&\times&f_M(x_i,x_j;p_i,p_j)
\end{eqnarray}
and for baryons can be re-written as
\begin{eqnarray}
\frac{dN_B}{d^2{\bf p}_{\rm T}}&=&g_B
\sum_{i\ne j\ne k}P_q(i)P_q(j)P_q(k)\nonumber\\
&\times&\delta^{(2)}({\bf p}_{\rm T}-{\bf p}_{i{\rm T}}-{\bf p}_{j{\rm T}}-
{\bf p}_{k{\rm T}})\nonumber\\
&\times&f_B(x_i,x_j,x_k;p_i,p_j,p_k)
\end{eqnarray}
Therefore, above, $P_q(i)$ and $P_{\bar q}(j)$ are probabilities carried by 
$i$-th test quark and $j$-th test antiquark with the condition that $\sum_{i}P_q(i) \delta^{(2)}({\bf p}_{\rm T}-{\bf p}_{i{\rm T}})=dN_q/d^2{\bf p}_{\rm T}$ and $\sum_{j}P_{\bar q}(j)\delta^{(2)}({\bf p}_{\rm T}-{\bf p}_{i{\rm T}})=dN_{\bar q}/d^2{\bf p}_{\rm T}$.
The advantage of this Monte-Carlo method is that it allows to 
treat the coalescence of high momentum particles with similar statistics as the one at low momentum. 

\section{Fragmentation}\label{section:Fragm}
The approach of hadronization discussed in this paper is based on a coalescence plus fragmentation 
modeling. 
As it has been clarified in Ref.s 
\cite{Fries:2003kq,Greco:2003mm,Fries:2008hs} at increasing $p_T$ the probability to coalescence 
decreases and eventually the standard independent fragmentation takes over.
In order to describe correctly the transition to the high momentum regime it is, therefore, necessary to include also the 
contribution from the fragmentation.
For the light quarks this is done by employing parton momentum distribution 
that at high $p_T>p_0\sim 3\,\rm \mbox{GeV}$ is evaluated in next-to-leading order (NLO) in a pQCD scheme. 
However in nucleus-nucleus collisions one must include also the modification due to
the jet quenching mechanism \cite{Gyulassy:2003mc,Wang:1998bha}. 
While the heavy-quark momentum spectra for both RHIC and LHC have been taken in accordance 
with the charm distribution in $p + p$ collisions within the Fixed Order + Next-to-Leading Log 
(FONLL), as given in Ref. \cite{Cacciari:2005rk,Cacciari:2012ny}.

We compute the coalescence probability $P_{coal}$ for each charm quark.
$P_{coal}$ is the probability that a charm quark with transverse momentum $p_T$
hadronize in a meson or a baryon according to the coalescence mechanism.
The overall normalization factor is determined by requiring the total recombination
probability for a charm to be 1 for a zero-momentum heavy quark.
This is done including the main  heavy flavor meson and baryon channels
listed in tables \ref{tab:D} and \ref{tab:Lambda}. \\
We notice that this choice, often considered by several groups, implies that the normalization factor $A_w$ in Eq. \ref{Eq:Wigner_M} and \ref{Eq:Wigner_B} is a factor of 2.4 larger than the value of 8 coming from the normalization to unity of the gaussian Wigner function. Such enhancement expected to take into account the enforcing of confinement by coalescence at vanishing momentum is nearly independent from the collisions energy, increasing by only $7\%$ at RHIC with respect to LHC.\\
From the coalescence probability we can assign a probability of fragmentation as
$P_{frag}(p_T)=1-P_{coal}(p_T)$ ($P_{frag}(p_T)=P_{c}(p_T)-P_{coal}(p_T)$). Therefore, the charm distribution function undergoing fragmentation, see Eq. (\ref{Eq:frag}), is evaluated convoluting the momentum distribution of heavy quarks which do not undergo to coalescence,
and is indicated as $dN_{fragm}/d^2p_Tdy$.

The hadron momentum spectra from the charm spectrum is given by:
\begin{equation}
\frac{dN_{had}}{d^{2}p_T\,dy}=\sum \int dz \frac{dN_{fragm}}{d^{2}p_T\, dy} \frac{D_{had/c}(z,Q^{2})}{z^{2}} 
\label{Eq:frag}
\end{equation}
where $z=p_{had}/p_{c}$ is the fraction of minijet momentum carried by the hadron and $Q^2=(p_{had}/2z)^2$ is the momentum scale 
for the fragmentation process. For $D$ and $\Lambda_{c}^{+}$ as $D_{had/c}(z,Q^{2})$ we employ the Peterson fragmentation function \cite{Peterson:1982ak}
\begin{eqnarray}
D_{had}(z) \propto 1/\bigg[ z \bigg[1-\frac{1}{z}-\frac{\epsilon_c}{1-z}\bigg]^2 \bigg]
\end{eqnarray}
where $\epsilon_c$ is a free parameter to fix the shape of the fragmentation function. 
For mesons the $\epsilon_c$ parameter is determined assuring that the experimental data 
on $D$ mesons production in $p+p$ collisions are well described by a fragmentation hadronization 
mechanism. The value it has been fixed to $\epsilon_c=0.006$ as discussed in \cite{Scardina:2017ipo}. 
In the absence of the $p+p$ data for the $\Lambda_c$ at RHIC and LHC energies, in this work we use 
the $e^{-}+e^{+}$ annihilation data to fix the shape of the fragmentation function
which gives an $\epsilon_c=0.02$ which is larger than the D meson as done in 
\cite{Das:2016llg}. The $\Lambda_c/D^0$ ratio is fixed to be about 0.1 in agreement with the
$e^++e^- $ analysis presented in \cite{Lisovyi:2015uqa}. However we will also explore the impact of higher values
of the ratio at the end of Section \ref{section:RHIC}.

\section{Fireball and parton distributions}\label{section:Fireball}
In this paper, we consider the systems created at RHIC in Au+Au collisions at $\sqrt{s_{NN}}=200$ \mbox{GeV}
and at LHC in Pb+Pb collisions at $\sqrt{s_{NN}}=2.76 \, \mbox{TeV}$.
Our coalescence approach is based on a fireball where the bulk of particles is a thermalized system
of gluons and $u,d,s$ quarks and anti-quarks. For the bulk properties we employ exactly the same
settings that have been already fixed to describe the spectra of light hadrons in Ref. \cite{Minissale:2015zwa}, 
where at $\tau=7.8 \, \mbox{fm}/c$, for LHC, and  $\tau=4.5 \, \mbox{fm}/c$, for RHIC, the system has a temperature of 
$T_{C}=165 \rm \, \mbox{MeV}$,
which is about the temperature for the cross-over transition in realistic lattice QCD calculation 
\cite{Borsanyi:2010cj}. \\
The longitudinal momentum distribution is assumed to be boost-invariant in the range $y\in(-0.5,+0.5)$.
To take into account for the quark-gluon plasma collective flow, we assume for the partons a radial flow profile as 
$\beta_T(r_T)=\beta_{max}\frac{r_T}{R}$, where $R$ is the transverse radius of the fireball. 
For partons at low transverse momentum,  $p_T<2 \,\mbox{GeV} $, hence we consider a thermal distribution
\begin{equation}
\label{quark-distr}
\frac{dN_{q,\bar{q}}}{d^{2}r_{T}\:d^{2}p_{T}} = \frac{g_{q,\bar{q}} \tau m_{T}}{(2\pi)^{3}} \exp \left(-\frac{\gamma_{T}(m_{T}-p_{T}\cdot \beta_{T} \mp \mu_{q})}{T} \right) 
\end{equation}
where $g_{q}=g_{\bar{q}}=6$ are the spin-color degeneracy of light
quarks and antiquarks, and the minus and plus signs are
for quarks and antiquarks, respectively. While $m_T$ is the transverse mass $m_T=\sqrt{p_T^2+m_{q,\bar{q}}^2}$.
For partons at high transverse momentum, $p_T> 2.5 \, \mbox{GeV}$, we consider the minijets that have undergone 
the jet quenching mechanism. Such a parton distribution can be obtained from pQCD calculations. 
As in Ref. \cite{Greco:2003mm} we have considered the initial $p_T$ distribution according to 
the pQCD and the thickness function of the Glauber model to go from pp collisions to
AA ones. Then we have quenched the spectra with the modelling as in Ref. \cite{Scardina:2010zz}
to reproduce the $p_T$ spectrum of pions
as observed experimentally at  $p_T \sim 8-10\,\rm \mbox{GeV}$.  
These parton spectra can be parametrized at RHIC as
\begin{equation}
\frac{dN_{jet}}{d^{2}p_{T}}=A \left( \frac{B}{B+p_{T}} \right)^{n}.
\end{equation}
These parametrization are the same used in \cite{Greco:2003xt,Greco:2003mm}
with the values given in the Table \ref{table1}, while in Table \ref{table2} are shown
the parameters used for the parametrization at LHC energies that is given by
\begin{equation}
\frac{dN_{jet}}{d^{2}p_{T}}=A_{1}\left[ 1+ \left( \frac{p_{T}}{A_{2}} \right)^{2}\right]^{-A_{3}}+A_{4}\left[ 1+ \left( \frac{p_{T}}{A_{5}} \right)^{2}\right]^{-A_{6}}.
\end{equation}
For heavy quarks we use the transverse momentum distribution obtained by solving the relativistic Boltzmann equation
\cite{Scardina:2017ipo} giving a good description of $R_{AA}$ and $v_{2}$ of $D$ mesons. They can be 
parametrized at RHIC and at LHC as

\begin{eqnarray}
\frac{dN_c}{d^2p_T}=
\left\{
\begin{array}{lr}
a_0 \exp{[-a_1 p_T^{a_2}]} & p_T\leq p_0 \\
a_0 \exp{[-a_1 p_T^{a_2}]}+a_3 \big(1+p_T^{a_4}\big)^{-a_5} & p_T\geq p_0 
\end{array}
\right. \nonumber
\end{eqnarray}
where $p_0=1.85 \, \mbox{GeV}$ and the parameters are given in Table \ref{tabCHARM}.
The number of heavy quark is estimated to be $dN_c/dy \simeq 2$ at RHIC and
$dN_c/dy \simeq 15$ at LHC, in agreement with the energy dependence of charm production cross section \cite{Adam:2016ich}. 
In the following calculation the charm quark mass used is $m_c = 1.3 \, \mbox{GeV}$.
\begin{table} [ht]
\begin{center}
\begin{tabular}{l c c c } 
& A$[1/\mbox{GeV}^{2}]$ & B[\mbox{GeV}] &n \\
\hline
\hline 
g & $3.18\cdot 10^{4}$  & 0.5 & 7.11 \\ 
\hline
$u,d$ & $9.79\cdot 10^{3}$ & 0.5 & 6.84 \\
\hline 
 $\bar{u},\bar{d}$ & $1.89\cdot 10^{4}$  & 0.5 & 7.59 \\ 
\hline 
$s$ & $6.51 \cdot 10^{3}$ & 0.5 & 7.36 \\ 
\hline
$\bar{s}$ & $8.02 \cdot 10^{3}$ & 0.5 & 7.57 \\ 
\hline 
\hline 
\end{tabular}
\end{center}
\caption{Parameters for minijet parton distributions at mid-rapidity from Au+Au collisions at $\sqrt{s_{NN}} = 200 \; \mbox{GeV}$}
\label{table1}
\end{table}

\begin{table} [ht]
\begin{center}
\begin{tabular}{l c c c c c c} 
& $A_{1}$ & $A_{2}$ & $A_{3}$ & $A_{4}$ & $A_{5}$ & $A_{6}$ \\
\hline
\hline 
$g$ & 23.46  & 4.84 & 8.08 & 2.78 & 2.79 & 2.31 \\ 
\hline
$u,d$ & 24.68  & 5.11 & 8.01 & 0.55 & 5.65 & 2.56 \\
\hline
$\bar u, \bar d$ & 23.12  & 5.05 & 8.21 & 0.57 & 5.62 & 2.58 \\
\hline
$s$  & 24.14  & 5.11 & 8.01 & 0.55 & 5.65 & 2.56 \\
\hline
$\bar s$ & 23.12  & 5.00 & 8.31 & 0.57 & 5.62 & 2.61 \\
\hline 
\hline
\end{tabular}
\end{center}
\caption{Parameters for minijet parton distributions at mid-rapidity from Pb+Pb collisions at $\sqrt{s_{NN}} = 2.76 \; \mbox{TeV}$}
\label{table2}
\end{table}

\begin{table} [ht]
\begin{center}
\begin{tabular}{l|ccccccc} 
\hline \hline
          RHIC  & $a_0$ & $a_1$ & $a_2$ & $a_3$  & $a_4$  & $a_5$   \\ 
 $p_T \leq p_0$  & 0.69 & 1.22  & 1.57  &        &        &        \\ 
 $p_T \geq p_0$  & 1.08 & 3.04  & 0.71  & 3.79   & 2.02   & 3.48  \\
\hline
          LHC   & $a_0$  & $a_1$ &  $a_2$ & $a_3$  & $a_4$  & $a_5$   \\ 
 $p_T \leq p_0$ & 1.97   & 0.35  &  2.47  &        &       &        \\ 
 $p_T \geq p_0$ & 7.95   & 3.49  &  3.59  & 87335  & 0.5   & 14.31  \\
\hline \hline
\end{tabular}
\end{center}
\caption{Parameters for charm distributions at mid-rapidity for Au+Au collisions at $\sqrt{s_{NN}} = 200 \; \mbox{GeV}$ and Pb+Pb collisions at $\sqrt{s_{NN}} = 2.76 \; A\mbox{TeV}$}
\label{tabCHARM}
\end{table}

\begin{figure}[t]
\centering
\includegraphics[width=\columnwidth, angle=0,clip]{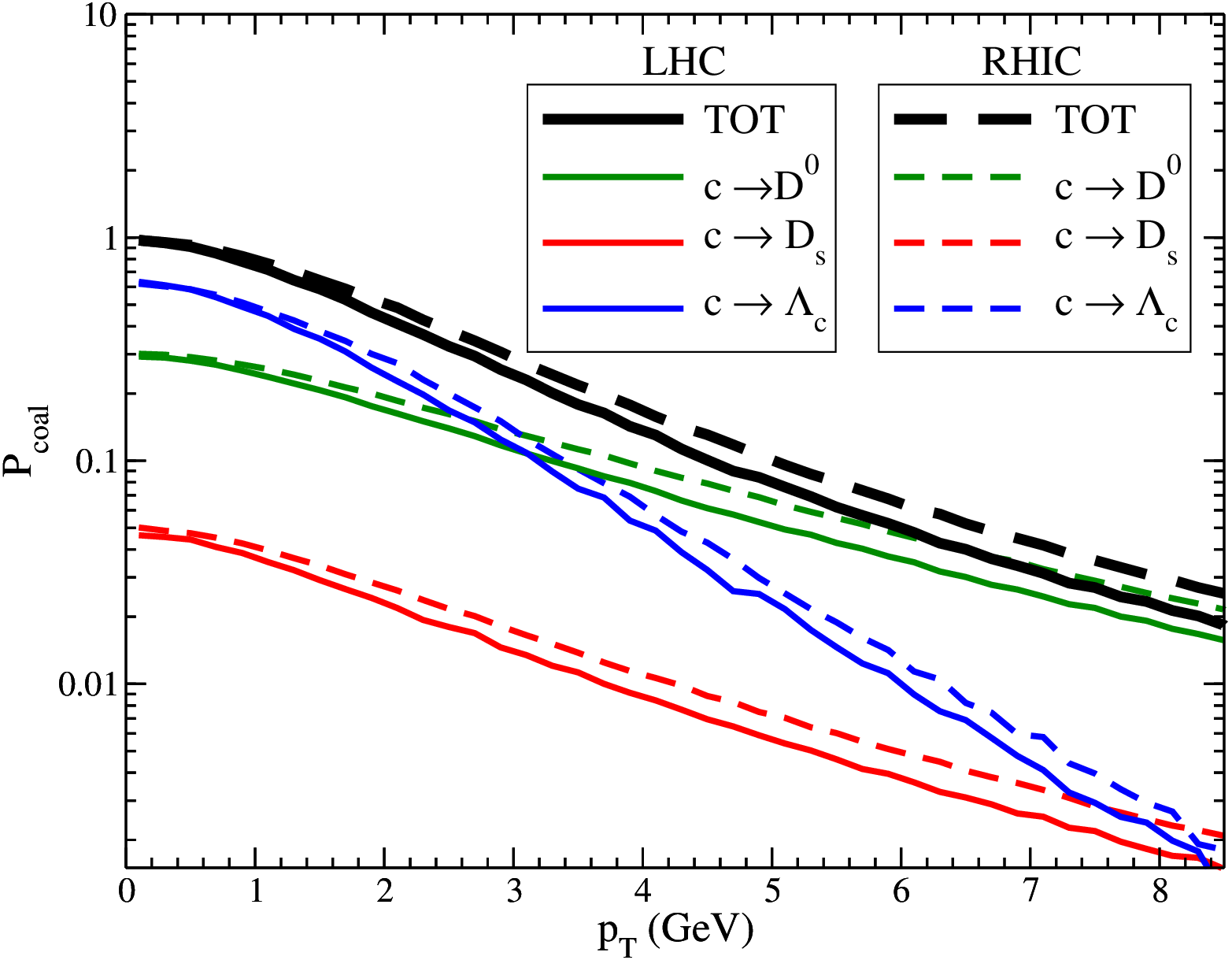}
\caption{\label{Fig:Pcoal} (Color online) The coalescence probabilities for charm quarks as a function of the 
transverse momentum. The solid lines refer to the case at LHC energies while dashed lines are for the case at RHIC energies.
Green, red and blue lines are the coalescence probabilities to produce $D^{0}$, $D_{s}$ and $\Lambda_C$ respectively. Black lines are the probabilities that a charm hadronize by coalescence in heavy meson or baryon.}
\end{figure}

Once that all the fireball parameters and widths have been set, it is possible to 
evaluate the coalescence probability.
In Fig. \ref{Fig:Pcoal} is shown the coalescence probabilities for charm quarks as
a function of the charm momentum, it is a decreasing function of $p_T$. 
This means that, at low momentum, charm quarks are more probable to hadronize
through coalescence with light partons from the thermalized medium, while at high $p_T$ 
the fragmentation becomes to be the dominant mechanism for charm hadronization.  
On the other hand, if an heavy quark is selected to fragment, based on the probability obtained subtracting the coalescence probability from the initial charm spectrum, its 
fragmentation is implemented by Eq. (\ref{Eq:frag}). 
Furthermore, has been considered that a charm quark has different fragmentation fraction into specific final charm hadron channels, as in Ref. \cite{Lisovyi:2015uqa}.

The comparison between the different coalescence probabilities in Fig. \ref{Fig:Pcoal} shows 
that a charm at high $p_T$ ($p_T \geq 3 \, \mbox{GeV}$) has a small probability to hadronize in $\Lambda_c$ 
by coalescence instead of recombine with a single light quark to form a $D^0$ meson. 
We notice that, at low momenta, having a coalescence probability for $\Lambda_c$ even larger than for $D^0$ is a quite peculiar
feature of the coalescence mechanism that we expect to lead to large values of the $\Lambda_c/D^0$ ratio,
as we will discuss in Sect. \ref{section:RHIC} and \ref{section:LHC}.

\section{The thermal model}\label{section:Thermal}
Measurements of the fragmentation of charm quarks into different hadrons
($D^0, D^+, D_s^+$ mesons and $\Lambda_c$) performed in deep
inelastic scattering in $e^{\pm}p$, $pp$ and $e^+e^-$ collisions
support the hypothesis that fragmentation is independent of the specific 
production process \cite{Lisovyi:2015uqa}.
Averages of the fragmentation fractions with a significantly reduced uncertainties 
have been obtained. These measurements lead to the particle ratios 
\begin{eqnarray}
\frac{D_s^{+}}{D^0}\bigg|_{pp,e^+e^-}\simeq 0.13 \;  ; \;
& \, \, &
\frac{\Lambda_c}{D^0}\bigg|_{pp,e^+e^-}\simeq 0.1 \nonumber
\end{eqnarray}

In the presence of the QGP medium a modification of the charm-quark hadronization is expected.
In the framework of the thermal or statistical hadronization models, 
the $p_T$-integrated ratios of D-meson abundances, were expected to be $D_s^+/D^0 \simeq 0.39$ 
which is larger by a factor of about three with respect to the values measured for $pp$ and $e^+e^-$ collisions.
While the estimated value for charmed baryon to meson ratio of about $\Lambda_c/D^0 \simeq 0.25$ 
\cite{Andronic:2003zv,Kuznetsova:2006bh,Andronic:2007zu}, is about a factor two larger with respect to $pp$ and $e^+e^-$ collisions according to Ref. \cite{Lisovyi:2015uqa}.

In a simplified version of the thermal model one assumes that, in relativistic heavy 
ion collisions, charmed and bottom hadrons are produced during hadronization 
of the quark-gluon plasma, and that they are both in thermal equilibrium at the phase transition temperature $T_C$. 
Therefore one can assume a thermal distribution. 
Assuming longitudinal boost invariance and neglecting the radial flow, 
the particle production at mid-rapidity approximately can be written as
\begin{equation}
\frac{dN_{H}}{dy \, p_T d p_T} = \frac{gV}{2\pi^2} m_{T} K_1(m_T/T_{C}),
\end{equation}\label{Eq:BW1}
where $K_n$ are the modified Bessel functions.
Furthermore by integrating with respect to the transverse momentum one can get the total 
number of heavy hadrons of mass $m$ inside a fireball of volume $V$, at 
temperature $T_{C}$, and per unit rapidity, which are given by
\begin{equation}
N_{H} = \frac{gV}{2\pi^2} m^2 T_{C} K_2(m/T_{C}),
\label{Eq.n-therm}
\end{equation}
where $g$ is the degeneracy of the particle, and $K_n$ are the modified Bessel
function. It is very well known that Eq. \ref{Eq.n-therm} would imply a large underestimate of  the charm
production because the initial abundance of charm is much larger than the chemical equilibrium value.
This leads to the inclusion in the thermal approach of a fugacity $\gamma_c$ factor \cite{Andronic:2003zv}.
However, in the following, we will discuss only the ratio of hadrons  with the presence of only one charm quark, so finally the ratio is independent on $\gamma_c$.\\
For $\Lambda_{c}/D^0$ using $m^{\Lambda_{c}^{+}}=2285 \, \mbox{MeV}$ and $m^{D^{0}}=1865 \, \mbox{MeV}$ 
the contribution coming from the ground state is given:
\begin{equation}
\frac{\Lambda_{c}^{+}}{D^{0}}\bigg|_0 = \frac{g_{\Lambda_{c}^{+}}}{g_{D^0}} \bigg( \frac{m^{\Lambda_{c}^{+}}}{m^{D^0}} \bigg)^{2} 
\frac{K_2(m^{\Lambda_{c}^{+}}/T_{C})}{K_2(m^{D^0}/T_{C})} \simeq 0.21
\end{equation}
with $g_{\Lambda_{c}^{+}}=2$ and $g_{D^{0}}=1$. 
The inclusion of the resonances have the effect to contribute significantly to the production of $\Lambda_{c}^{+}$ and $D^0$ 
while the final value of the $\Lambda_{c}^{+}/D^0$ ratio is only enhanced by about a $20\%$, as we show in the
following. In fact the main contribution
to $D^{0}$ comes from $D^{* +}$ and $D^{* 0}$ according to the decays listed in Table \ref{tab:D} and this gives an enhancement 
to $D^0$ of about $D^{*+}(2007)/D^0 \simeq 1.68 \cdot 1.401 \simeq 2.35$ where we have included the corresponding branching ratios.
While for $\Lambda_{c}^{+}$ the main contribution comes from 
$\Sigma_{c}^{+}(2625)$, $\Sigma_{c}^{+}(2455)$ and $\Lambda_{c}^{*+}(2625)$ with the decays shown in Table \ref{tab:Lambda}. 
Therefore the contribution to $\Lambda_{c}$ are 
$\Sigma_{c}^{*}(2625)/\Lambda_{c}^{+} \simeq 1.65$, $\Sigma_{c}^{*}(2455)/\Lambda_{c}^{+} \simeq 1.182$ and
$\Lambda_{c}^{*+}(2625)/\Lambda_{c}^{+} \simeq 0.38$ respectively.
Thus the final ratio is approximately given by
\begin{eqnarray}
&&\frac{\Lambda_{c}^{+}}{D^{0}}=\frac{\Lambda_{c}^{+}}{D^{0}}\bigg|_0 \times \nonumber \\
&\times& \frac{1+\Sigma_{c}^{*}(2625)/\Lambda_{c}^{+}+\Sigma_{c}^{*}(2455)/\Lambda_{c}^{+}+\Lambda_{c}^{*+}/\Lambda_{c}^{+}}{1+(D^{*+}/D^0)}\simeq 0.26 \nonumber
\end{eqnarray}
This simple calculation shows that in a thermal model an
enhancement of the baryon-to-meson ratio by a factor 2 is expected with respect to fragmentation and 
this simple estimation is in agreement with more sophisticated calculation within the SHM \cite{Andronic:2007zu,Andronic:2010dt,Kuznetsova:2006bh}. 

The $p_T$ dependence of the baryon-to-meson ratio can be evaluated easily from this blast-wave model. In fact
\begin{equation}
\frac{\Lambda_{c}^{+}}{D^{0}}=
\frac{g_{\Lambda_{c}^{+}}}{g_{D^0}} 
\frac{m^{\Lambda_{c}^{+}}_{T}}{m^{D^0}_{T}} 
\frac{K_1(m^{\Lambda_{c}^{+}}_{T}/T)}{K_1(m^{D^0}_{T}/T)}
\end{equation}
with $m_T=\sqrt{m^2+p_T^2}$. This ratio is an increasing function with the transverse momentum and for very large transverse momentum it saturates to the relative ratios of the degeneracy $\Lambda_C/D^0 \to g_{\Lambda_{c}^{+}}/g_{D^0}=2$, see also Fig. \ref{Fig:LambdaD0_ratio}.
Furthermore for low hadron transverse momentum we have 

\begin{eqnarray}
\label{Eq:thermal_ratio}
&& \frac{\Lambda_{c}^{+}}{D^{0}}\bigg|_{p_T \simeq 0}=\frac{g_{\Lambda_{c}^{+}}}{g_{D^0}} \frac{m^{\Lambda_{c}^{+}}}{m^{D^0}} \frac{K_1(m^{\Lambda_{c}^{+}}/T)}{K_1(m^{D^0}/T)} \simeq \nonumber \\
&&\simeq \frac{g_{\Lambda_{c}^{+}}}{g_{D^0}} \bigg( \frac{m^{\Lambda_{c}^{+}}}{m^{D^0}} \bigg)^{1/2} e^{-(m^{\Lambda_{c}^{+}}-m^{D^0})/T_C}\simeq 0.17
\end{eqnarray}
This shows that, in general, within the blast-wave description, the baryon-to-meson ratio is 
exponentially suppressed with the mass of the hadrons.

In the rest of the paper we will analyze the production of charmed hadron within a coalescence plus 
fragmentation model.
In our calculations, all major hadron channels have been incorporated,
including the first excited states for $D$ mesons and $\Lambda_C$, as discussed in this Section.
For the resonances the coalescence probability is multiplied by a suppression factor
that takes into account for the Boltzmann probability to populate an excited state of energy 
$E + \Delta \, E$, at a temperature T. In particular the coalescence probability for excited states is augmented
by the statistical model factor $(m_{H^*}/m_{H})^{3/2}$ $\times \exp{(-\Delta E/T)}$
with $\Delta \,E=E_{H^*}-E_H$ and $E_{H^*}=(p_T^2+m_{H^*}^2)^{1/2}$ and
$m_{H^*}$ is the mass of the resonance. Of course also the degeneracy factors that come from the different 
values of isospin and total angular momentum are taken in account.

\begin{table} [ht]
\begin{center}
\begin{tabular}{cccccc}
\hline
Meson & Mass (MeV) & I (J) & \\
\hline 
$D^+ =\bar{d}c$		& 1869 & $\frac{1}{2} \,(0)$	&\\
$D^0 =\bar{u}c$ 	& 1865 & $\frac{1}{2} \,(0)$	&\\
$D_{s}^{+} =\bar{s}c$	& 2011 & $0 \,(0)$		& \\
\hline
Resonances & & & Decay modes& B.R. \\
\hline
$D^{*+} =\bar{d}c$	& 2010 & $\frac{1}{2} \, (1)$	& $D^0 \pi^+$ & $68\%$ \\
&&& $D^+ X$& $32\%$ \\
$D^{*0} =\bar{u}c$	& 2007 & $\frac{1}{2} \, (1)$	& $D^0 \pi^0$ & $62\%$ \\
&&& $D^0 \gamma$& $38\%$ \\
$D_{s}^{*+} =\bar{s}c$	& 2112 & $0 \, (1)$		& $D_{s}^+ X$ & $100\%$ \\
\hline
\end{tabular}
\end{center}
\caption{
Charmed mesons considered in this work. Top section the ground states considered 
while in the bottom section the first exited states including their decay modes with 
their corresponding branching ratios as given in Particle Data Group \cite{Agashe:2014kda}.
\label{tab:D}}
\end{table}
\begin{table} [ht]
\begin{center}
\begin{tabular}{cccccc}
\hline
Baryon & Mass (MeV) & I (J) & \\
\hline 
$\Lambda_c^+ =udc$	& 2286 & $0 \, (\frac{1}{2})$	&\\
\hline
Resonances & & & Decay modes& B.R. \\
\hline
$\Lambda_c^+ =udc$	& 2595 & $0 \, (\frac{1}{2})$	& & \\
$\Lambda_c^+ =udc$	& 2625 & $0 \, (\frac{3}{2})$	& & \\
$\Sigma_c^+ =udc$	& 2455 & $1 \, (\frac{1}{2})$	&$\Lambda_c^+ \pi$ & $100\%$\\
$\Sigma_c^+ =udc$	& 2520 & $1 \, (\frac{3}{2})$	&$\Lambda_c^+ \pi$ & $100\%$\\
\hline
\end{tabular}
\end{center}
\caption{
Charmed baryons considered in this work. Top section the ground states considered 
while in the bottom section the first exited states including their branching 
ratios as given in Particle Data Group \cite{Agashe:2014kda}.
\label{tab:Lambda}}
\end{table}

\section{Heavy Hadron transverse momentum spectra and ratio at RHIC}\label{section:RHIC}
In this section, we show results for the transverse momentum spectra
of $D^0$, $D^+$, $D_s$ mesons and for $\Lambda_c$ using the model described in 
previous sections for $Au+Au$ collisions at $\sqrt{s}=200 \, \rm \mbox{GeV}$ in central collisions. For the coalescence 
contribution the effects due to gluons in the quark-gluon plasma is taken into account 
by converting them to quarks and anti-quark pairs according to the flavour compositions in
the quark-gluon plasma, as assumed in \cite{Biro:1994mp,Greco:2003mm}.   
We include ground state hadrons as well as the first excited resonances listed in tables \ref{tab:D} and \ref{tab:Lambda}.

\begin{figure}[t]
\centering
\includegraphics[width=\columnwidth, angle=0,clip]{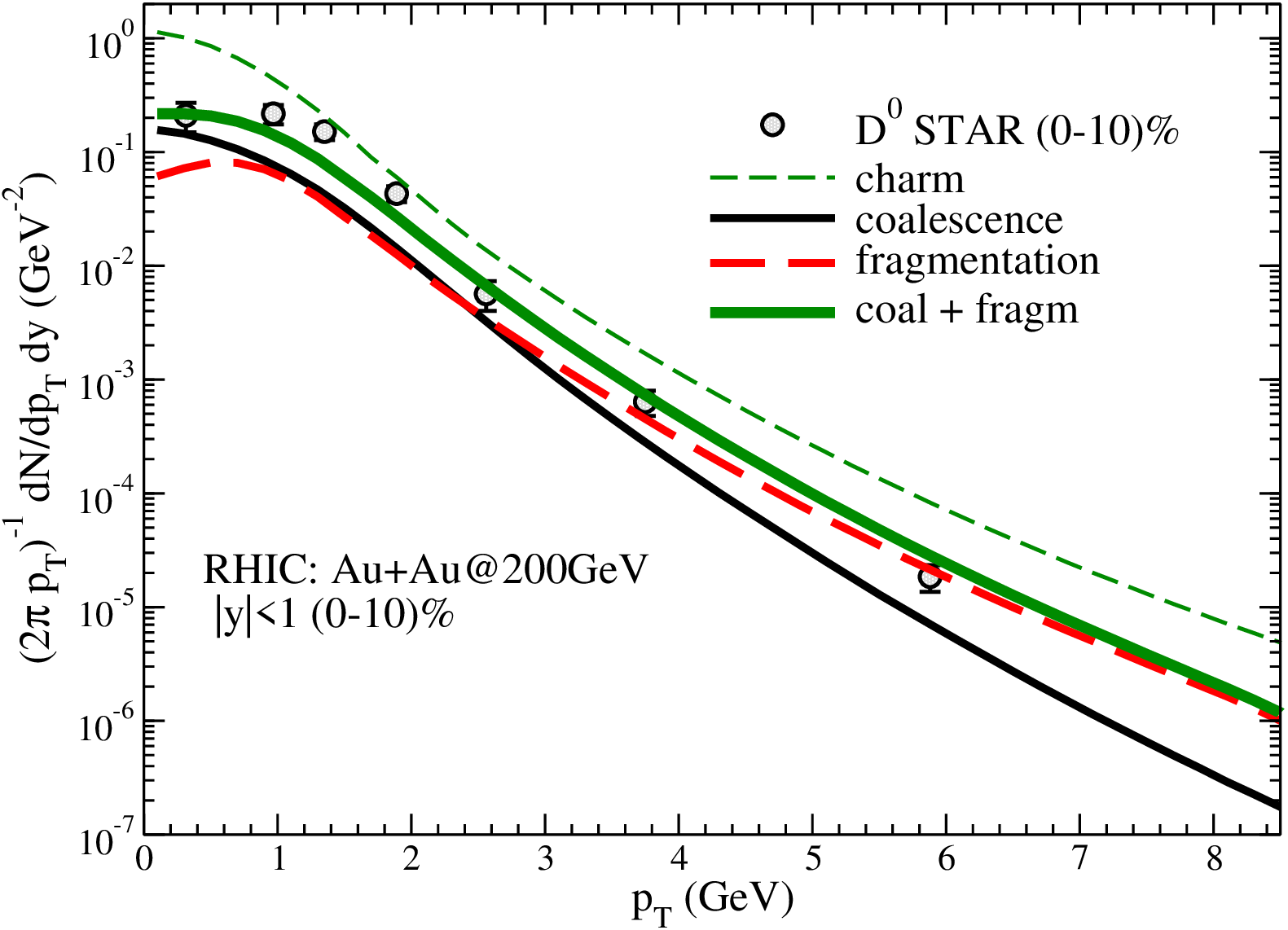}
\caption{
\label{Fig:D0_RHIC} 
(Color online) Transverse momentum spectra for $D^0$ meson at mid-rapidity 
for $Au+Au$ collisions at $\sqrt{s}=200 \, \mbox{GeV}$ and for $(0-10\%)$ centrality.
Green dashed line refers to the charm spectrum. Black solid and red dashed lines refer to the $D^0$
spectrum from only coalescence and only fragmentation respectively while green solid line refers 
to the sum of fragmentation and coalescence processes. 
Experimental data taken from \cite{Adamczyk:2014uip}.}
\end{figure}
In Fig. \ref{Fig:D0_RHIC} we show the $p_T$ spectra for $Au+Au$ collisions at mid-rapidity
for $(0-10\%)$ centrality. The thin green dashed line refers to the $p_T$ spectrum of charm quarks, 
while the black solid line and the red dashed line refer to the spectra of $D^0$ meson obtained by 
the contribution from pure coalescence and fragmentation respectively. Moreover, we can see that 
the contribution of both mechanism is about similar for $p_T \lesssim 3 \, \rm \mbox{GeV}$ and at higher $p_T$ the 
fragmentation becomes the dominant hadronization mechanism. Finally, the green solid line is the 
contribution of both coalescence and fragmentation and, as shown, both hadronization mechanism 
are needed to have a good description of the experimental data, especially at $p_T < 4 \,\rm \mbox{GeV}$. 

\begin{figure}[t]
\centering
\includegraphics[width=\columnwidth, angle=0,clip]{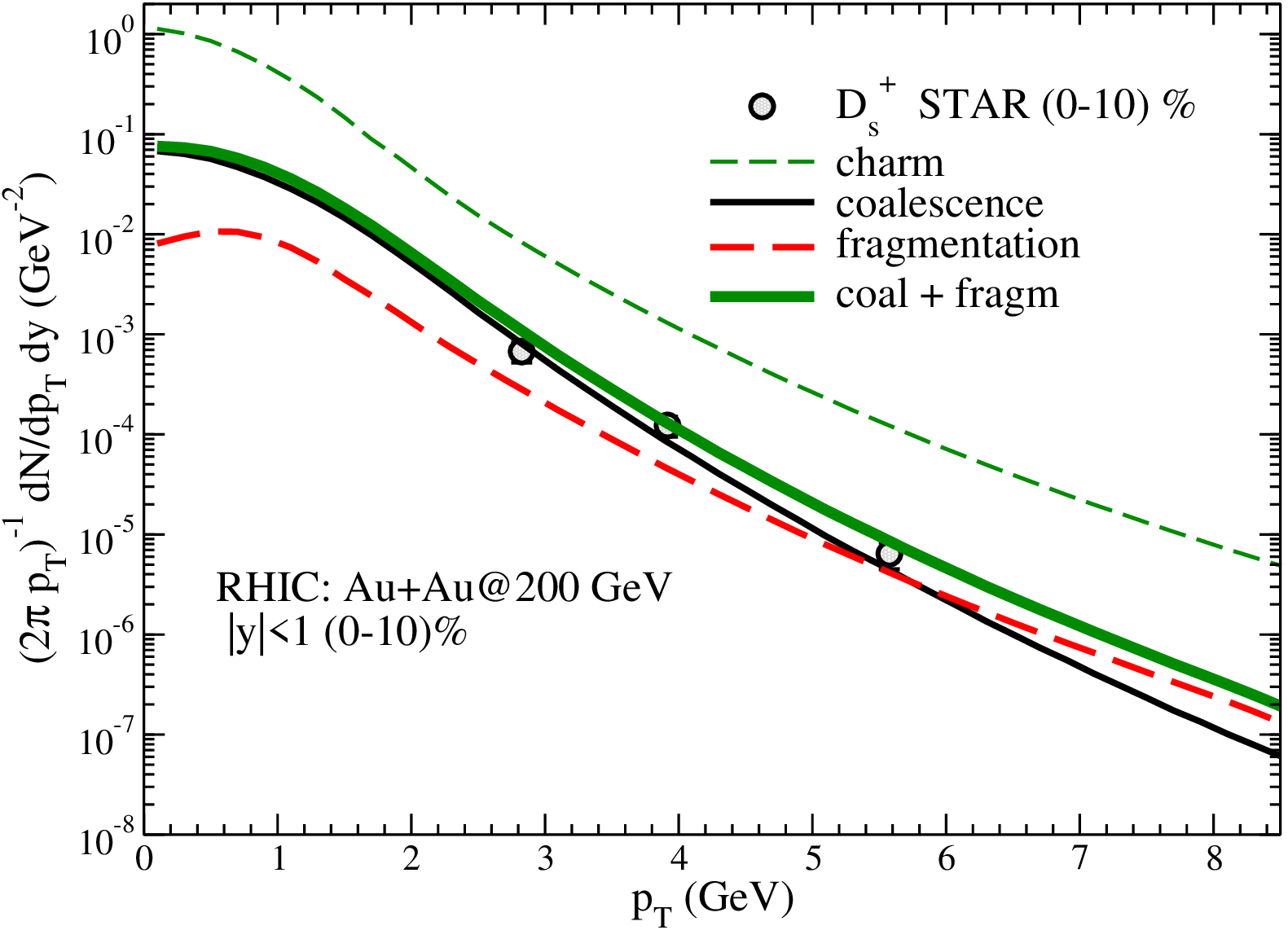}
\caption{
\label{Fig:Ds_RHIC}
(Color online) Transverse momentum spectra for $D_{s}^{+}$ meson at mid-rapidity 
for $Au+Au$ collisions at $\sqrt{s}=200 \, \mbox{GeV}$ and for $(0-10\%)$ centrality.
Green dashed line refers to the charm spectrum. Black solid and red dashed lines refer to the $D_{s}^{+}$
spectrum from only coalescence and only fragmentation respectively while green solid line refers 
to the sum of fragmentation and coalescence processes. 
Experimental data taken from \cite{Zhou:2017ikn}.
}
\end{figure}

In Fig. \ref{Fig:Ds_RHIC} one shows the transverse momentum spectra for the $D_s^+$ meson
at mid-rapidity at RHIC energies $\sqrt{s}=200 \, \mbox{GeV}$ and for $(0-10\%)$ centrality.
Comparing the relative contributions by coalescence and fragmentation to the 
production of $D_s^+$, black solid and red dashed lines respectively, we observe that at low $p_T$ 
coalescence is the dominant mechanism, while fragmentation play a significant role at $p_T \gtrsim 4 \rm \mbox{GeV}$.
This is related to the fact that the fragmentation fraction for $D_s^+$ is quite small, about $8 \%$
of the total heavy hadrons produced, according to Ref. \cite{Lisovyi:2015uqa}. Again the comparison with the experimental 
data shows that only the inclusion of both hadronization mechanisms provide a quite good prediction.
Furthermore we expect that coalescence leads to an enhancement of the $D_s^+$ production;
a feature that seems to be present in first experimental data on $R_{AA}$ at ALICE and predicted
in \cite{He:2012df}.  The different relative contribution of coalescence and fragmentation for $D_s$ w.r.t. $D^0$ leads
to an enhancement of the ratio $D_s/D^0$ of about 0.3 in the wide region were coalescence dominates, $p_T \lesssim 4 \,\rm \mbox{GeV}$.

\begin{figure}[t]
\centering
\includegraphics[width=\columnwidth, angle=0,clip]{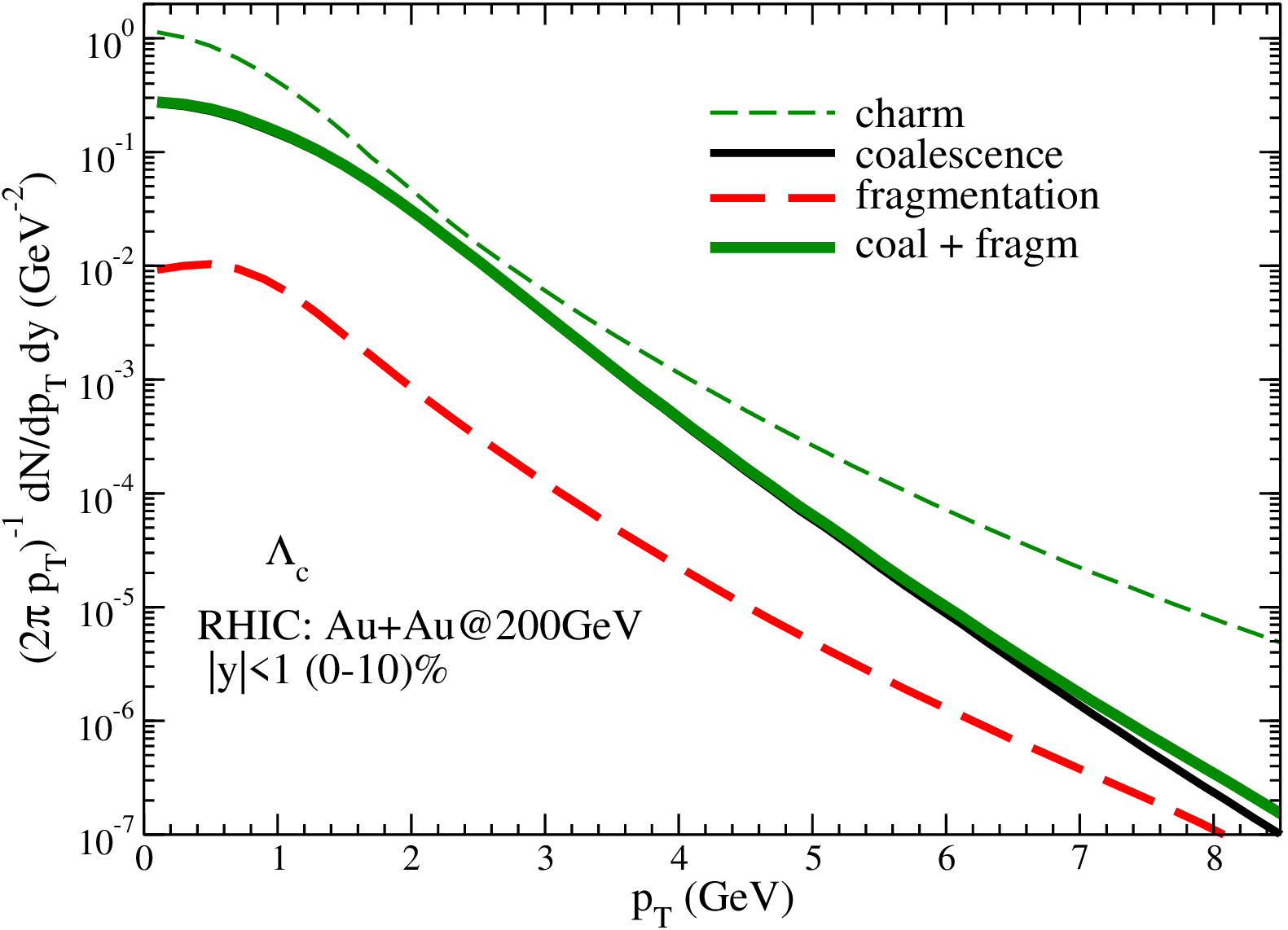}
\caption{\label{Fig:LambdaC_RHIC}
(Colour online) 
Transverse momentum spectra for $\Lambda_{c}^{+}$ baryon at mid-rapidity 
for $Au+Au$ collisions at $\sqrt{s}=200 \, \mbox{GeV}$ and for $(0-10\%)$ centrality.
Green dashed line refers to the charm spectrum. Black solid and red dashed lines refer to the $\Lambda_c^{+}$
spectrum from only coalescence and only fragmentation respectively while green solid line refers 
to the sum of fragmentation and coalescence processes. 
}
\end{figure}

For $\Lambda_{c}^{+}$ baryon we have included main hadronic channels including the 
ground state and first excited states. The main resonances contribution comes from 
$\Sigma_{c}^{*}(2520)$ and $\Sigma_{c}(2455)$ that decay almost $100\%$ in $\Lambda_{c}^{+}$
via the decays $\Sigma_{c}^{*} \rightarrow \Lambda_{c}^{+} \pi$ and 
$\Sigma_{c} \rightarrow \Lambda_{c}^{+} \pi$.
In Fig. \ref{Fig:LambdaC_RHIC} we show the $\Lambda_{c}^{+}$
transverse momentum spectrum at mid-rapidity and RHIC energies
for $0-10\%$ centrality, including coalescence and fragmentation 
by solid and dashed lines respectively.
We notice that the coalescence mechanism is the dominant mechanism for 
the $\Lambda_{c}^{+}$ production for $p_T \lesssim 7 \, \mbox{GeV}$.
This is due to the combination of two conditions:
on one hand it is related to the the ratio for $\Lambda_c^+/D^0$ in the 
fragmentation analysis of Ref. \cite{Lisovyi:2015uqa} that is very small, because the fragmentation fraction in $\Lambda_c^+$ is about the $6 \%$ of the total produced heavy hadrons.
On the other hand, as known for light hadrons, the coalescence contribution is more 
important for baryons with respect to mesons \cite{Minissale:2015zwa}, essentially because the mechanism is not based on the production
of two quarks from the QCD vacuum, but uses quarks that are already present 
abundantly in the QGP bulk.
Here, the result is an enhancement of about an order of magnitude for the $\Lambda_{c}^{+}$
production. We have also to mention that, for this result, it is important to normalize coalescence 
in such a way that for $p \to 0$ all charm hadronize by coalescence. Using a standard normalization 
as in \cite{Minissale:2015zwa} the dominance of coalescence is still present but the yield of $\Lambda_{c}^{+}$ will be 
reduced by about a factor of $5-6$.
\begin{figure}[t]
\centering
\includegraphics[width=\columnwidth, angle=0,clip]{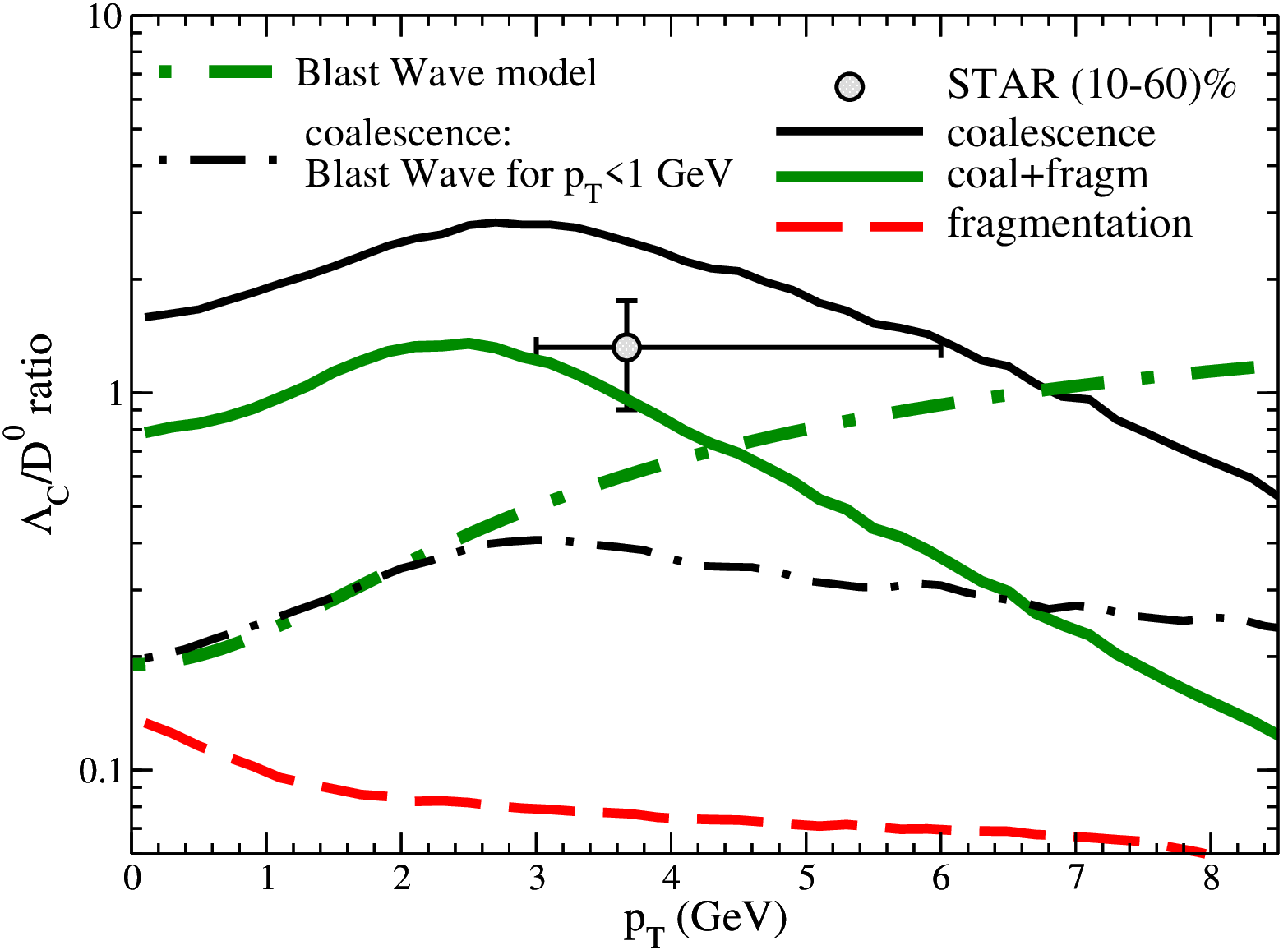}
\caption{\label{Fig:LambdaD0_ratio} 
(Color online) 
$\Lambda_{c}^{+}$ to $D^0$ ratio as a function of $p_T$ and at mid-rapidity for $Au+Au$ collisions at $\sqrt{s}=200 \, \mbox{GeV}$ 
and for $(10-60\%)$ centrality. Experimental data taken from \cite{Zhou:2017ikn}. 
Black solid and red dashed lines refer to the case from only coalescence and only fragmentation respectively 
while green solid line refers to the sum of fragmentation and coalescence processes.
Finally, the dot-dashed lines refer to the blast wave model including the effect of radial flow(green) and coalescence with wave function widths $\sigma_{j}$ 
of $D^0$ and $\Lambda_c^{+}$ changed to have the thermal ratio at $p_T \rightarrow 0$ (black), see the text for more details.
}
\end{figure}

The coalescence mechanism is naturally able to predict the  
baryon/meson enhancement for light flavour at intermediate 
transverse momentum, with a quite well description of the rise 
at low $p_T$ up to the peak region at $p_T \simeq 3 \, \mbox{GeV}$ 
and then the falling-down behaviour \cite{Minissale:2015zwa}.
Moreover is able to describe naturally the region of $p_{T}= 2 -4 \mbox{GeV}$, 
which is the region where this ratios $p/\pi^{+}$, $p/\pi^{-}$ and 
$\Lambda/2 K_{s}$ reach a value close to the unity, 
which is a stronger enhancement with respect to the 
one observed in $pp$ collisions.
In Fig. \ref{Fig:LambdaD0_ratio} we show the results for the 
$\Lambda_{c}^{+}/D^{0}$ ratio in comparison with the STAR experimental 
data shown by circle. Solid black line is the result obtained by 
pure coalescence, while the red dashed line is the case with pure fragmentation according to \cite{Lisovyi:2015uqa}. 
As shown by comparing red dashed line and black solid line,  
the coalescence by itself predicts a rise and fall of the baryon/meson ratio.
The inclusion of fragmentation reduces the ratio, and 
we can see that in the peak region a quite good agreement with the
only experimental data by STAR is reached (green solid line). 
Notice that in our calculation we obtain similar baryon/meson ratio to the one predicted in \cite{Oh:2009zj}.
However we note that compared with measured light baryon/meson ratios like $\bar{p}/\pi^-$ and 
$\Lambda/K_S^0$ ratios (see \cite{Adler:2003cb,Adler:2004zd,Abelev:2014laa}), the obtained
$\Lambda_{c}^{+}/D^0$ ratios has a different behaviour.
This heavy baryon/meson ratio is thus much flatter than the light baryon/meson ratios.
In fact for $p_T \rightarrow 0$ hadronization by coalescence and 
fragmentation predict $\Lambda_{c}^{+}/D^0 \simeq 0.75$ which is much larger 
with respect to the one measured or calculated by coalescence for light baryon/meson ratio, with
$\Lambda/K^0 \simeq 0.1$ for $p_T \rightarrow 0$ \cite{Fries:2008hs,Minissale:2015zwa}.
This behaviour comes from the large mass of heavy quarks.
In fact, in the non relativistic limit, as shown in \ref{App:coal}, an approximate solution of 
the coalescence integral predicts that the baryon-to-meson ratio is proportional to the reduced 
mass  $\mu_2$ of the baryon, that for a cqq system is about a factor 3 larger than for a qqq system
see Eq. (\ref{Eq.App.mu2}). 

It is interesting to compare this ratio with the one obtained by a blast wave model, where
the ratio is given by the thermal spectrum 
of $\Lambda_c$ with a thermal spectrum of $D^0$ , including the radial flow $\beta_{T}(r_{T})$ and the resonance 
decays taken into account in the coalescence calculation (see Table \ref{tab:D} and \ref{tab:Lambda}).
The result is shown in Fig. \ref{Fig:LambdaD0_ratio} by green dot-dashed line. We can see that, within this simplified blast wave model, 
the ratio is an increasing function of the transverse momentum and, in particular, in the limit for $p_T \rightarrow 0$ has a value of about $0.2$; consistent 
with the average value given by more sophisticated thermal models \cite{Andronic:2003zv}.

The low momentum region of the heavy baryon-to-meson ratio is interesting because coalescence models and thermal 
models predict a quite different trend. In fact as shown in Eq.\ref{Eq:thermal_ratio} the thermal model gives a small value, due to the exponential suppression with respect to the baryon mass. With our coalescence model
there is a quite milder $p_T$ dependence, because the gain in momentum of an additional light quark in $\Lambda_{c}^{+}$
is quite small.
In fact it is true that one can predict a peak in the $\Lambda_c/D^0 \approx \,1$, but this
is not associated to a small value of the ratio as $p_T \rightarrow 0$, at variance with the ratio
of light hadrons like $p/\pi$ and $\Lambda/K^0$, as predicted in \cite{Fries:2008hs,Minissale:2015zwa}.
Therefore the study $\Lambda_{c}/D^0$ ratio is a good tool 
to disentangle different hadronization mechanisms once the data will be available mostly in the low $p_T$ regime.

This can be further seen by the black dot-dashed line in Fig. \ref{Fig:LambdaD0_ratio}, which is the calculation 
within coalescence plus fragmentation where the Wigner function widths have been fixed in order to reproduce the 
thermal model at low transverse momentum (green dot-dashed line). This is the strategy behind the predictions
in Ref. \cite{Ghosh:2014oia} based on the same coalescence model but with the idea that
in the low $p_T$ regime the SHM leads to a correct prediction of the ratio.
Within the coalescence model it is possible to do this by choosing a different value for the $\omega$ parameter that defines the width of the wave function.
Choosing for $D$ meson $\sigma_p=\sigma_r^{-1}=0.2 \,\rm \mbox{GeV}$, and for $\Lambda_c$ the values 
$\sigma_1=0.326 \,\rm \mbox{GeV}$ and  $\sigma_2=0.63 \,\rm \mbox{GeV}$; that give respectively a factor 1.4 larger for the root mean 
square charge radius of the $D^0$  w.r.t. the quark model,
and a factor 2 smaller for the $\Lambda_c$. At the same time, this change results in a smaller value for the $\Lambda_c^+/D^0$ ratio.
However, we can see, that if we tune coalescence to agree with a thermal approach then we cannot reach values of the ratio close to 1 
\cite{Das:2016llg}. Therefore it seems that, anyway, in a coalescence model for charmed hadrons one cannot have a peak
of $\Lambda_c/D^0 \gtrsim \,1$ and a $\Lambda_c/D^0 \gtrsim \,0.2$ in the low $p_T$ region.

\begin{figure}[t]
\centering
\includegraphics[width=\columnwidth, angle=0,clip]{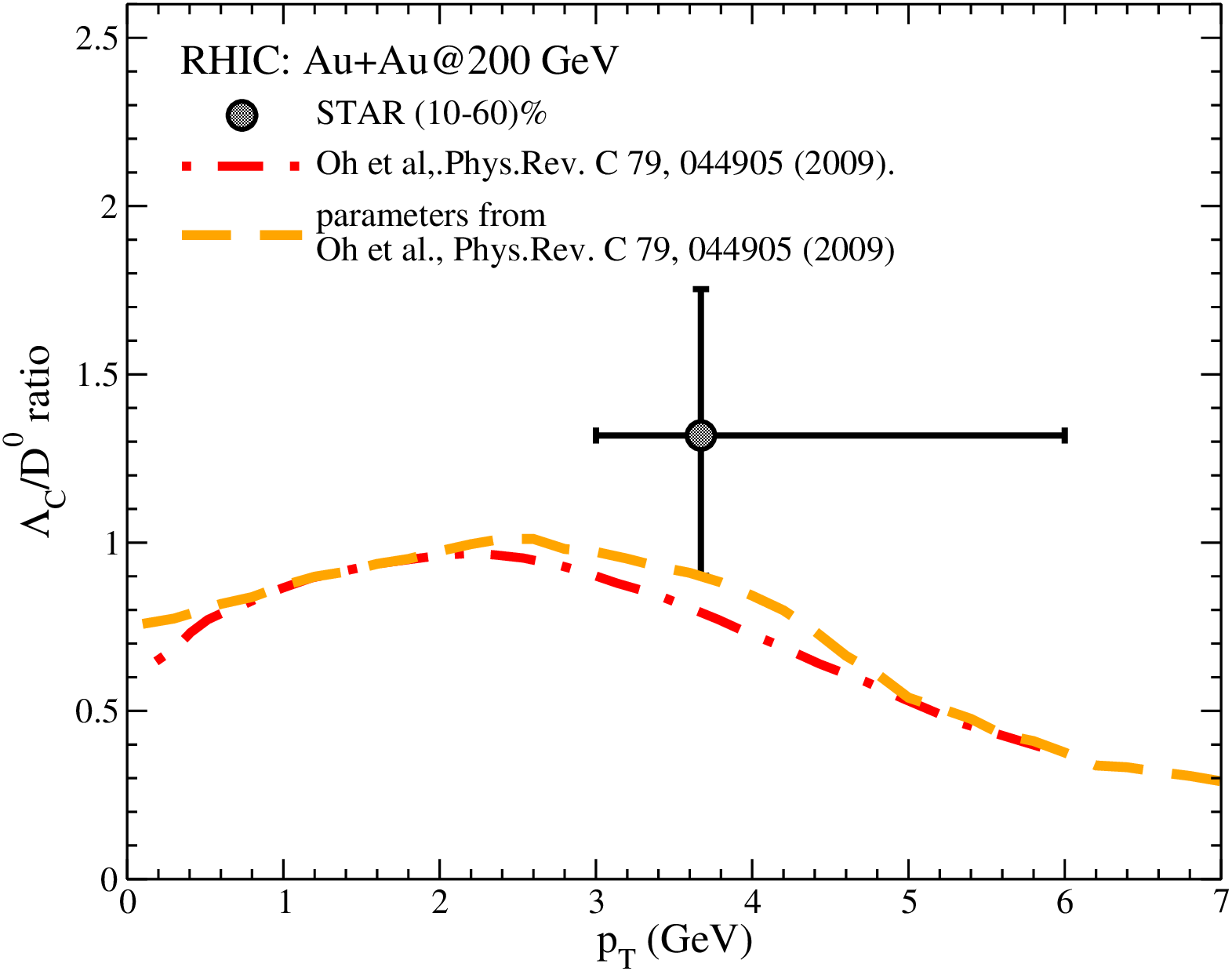}
\caption{\label{Fig:comparison1}
(Color online) $\Lambda_{c}^{+}/D^0$ ratio as a function of $p_T$ 
and at mid-rapidity for $Au+Au$ collisions at $\sqrt{s}=200 \, \mbox{GeV}$. These calculations are 
obtained including only coalescence. Orange dashed line refer to the case where the fireball parameters and 
widths have been fixed like in \cite{Oh:2009zj}.
The red dot-dashed line refer to the calculation in \cite{Oh:2009zj}.
}
\end{figure}

As mentioned in the introduction an early prediction with a peak in $\Lambda_c/D^0$ of about  one or even larger
was presented in Ref. \cite{Oh:2009zj}. Later in Ref. \cite{Ghosh:2014oia} with a coalescence plus fragmentation
model a quite smaller ratio was predicted, that indeed corresponds to the black dot-dashed line in Fig. \ref{Fig:LambdaD0_ratio}.
We want to clarify that the formulation of the coalescence process in our approach and the one in Ref.  \cite{Oh:2009zj} is practically identical and the differences are
due to the mean square radius assumed for $\Lambda_c$ and $D^0$.
To better clarify this point in a transparent way, we have set the fireball parameters like the temperature T and the radial flow
as in Ref. \cite{Oh:2009zj}, i.e. $T=200\,\rm \mbox{MeV}$ and $\beta_T=0$. Furthermore in our approach we set the underlying $\omega$
parameter, determining the width of the Wigner function equal to $0.537 \, \rm \mbox{fm}^{-1}$ for mesons and baryons, again as done
in Ref. \cite{Oh:2009zj}; with the justification that this choice of the Wigner function widths, in the $p_T \rightarrow 0$ limit, gives an hadronization by coalescence for all the charm quark.
This leads, respectively, to charge radii for charmed hadrons
of 0.74 fm for $D^{0}$ meson and 0.78 fm for $\Lambda_c^{+}$ baryon; 
which are factors of 1.77 and 1.95 larger than those predicted by quark models
in Ref.s \cite{Hwang:2001th,Albertus:2003sx}.
In Fig. \ref{Fig:comparison1}, we show by
 orange dashed line the result with our code in comparison with the results of Ref. \cite{Oh:2009zj},
 shown by the dot-dashed line.
We can see that the results are very similar, in fact technically the difference is only that in \cite{Oh:2009zj} a non-relativistic
approximations is employed, while in our case we solve, by Monte Carlo methods, the full integral in Eq. \ref{eq-coal}.
As we can see, this difference does not lead to any significant dissimilarity in the final outcome.
We note that even if the results in Fig. \ref{Fig:comparison1} show a peak in the $\Lambda_c/D^0$ ratio of about 1, this is obtained without including the fragmentation and therefore, at finite $p_T$, there are charm quarks that do not
undergo hadronization.

A main novelty of the present work is the inclusion of both hadronization mechanism, ensuring that all charm quarks hadronize also at 
finite $p_T$ and the employment of a width for the hadron Wigner function consistent with the quark model  \cite{Hwang:2001th,Albertus:2003sx}.

\section{Heavy hadron spectra and ratio at LHC}\label{section:LHC}
In this section we show the results from coalescence plus fragmentation 
in comparison to the recent experimental data from $Pb+Pb$ collisions
at $\sqrt{s}=2.76 \, \mbox{TeV}$.
We mention that the results have been obtained without any change 
or addition of microscopic parameters $\sigma_{j}$ with respect to 
the one at RHIC in the previous Section. 
The only parameters that have been changed are the ones 
related to the dimension of the fireball,
in particular, the radial flow and volume of the hadronizing fireball that,
as described in the previous section, have been constrained
by the total transverse energy and multiplicity at LHC;
and have the same values used for light hadron calculations in Ref. \cite{Minissale:2015zwa}.

\begin{figure}[t]
\centering
\includegraphics[width=\columnwidth, angle=0,clip]{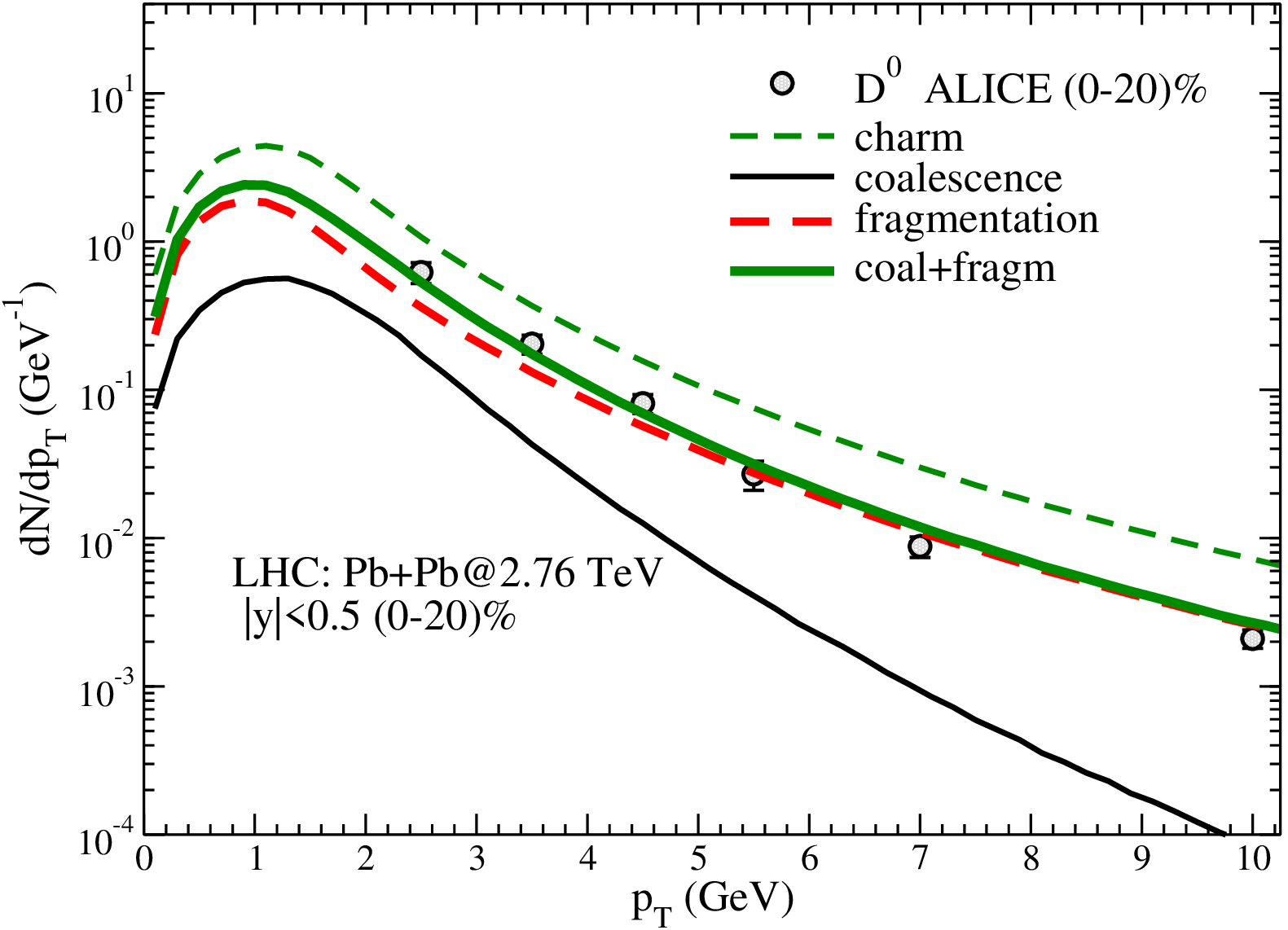}
\caption{
\label{Fig:D0_LHC} 
(Color online) 
Transverse momentum spectra for $D^0$ meson at mid-rapidity 
for $Pb+Pb$ collisions at $\sqrt{s}=2.76 \, \mbox{TeV}$ and for $(0-20\%)$ centrality.
Green dashed line refers to the charm spectrum. Black solid and red dashed lines refer to the $D^0$
spectrum from only coalescence and only fragmentation respectively while green solid line refers 
to the sum of fragmentation and coalescence processes. 
Experimental data taken from \cite{ALICE:2012ab}.}
\end{figure}
In Fig. \ref{Fig:D0_LHC} is shown the transverse momentum spectrum for 
$D^{0}$ meson at mid-rapidity for $(0-20\%)$ centrality. 
The total $D^0$ spectrum (coalescence plus fragmentation) shown by 
green line is in a good agreement with the experimental data. 
The black solid and red dashed lines refer to the contribution for pure coalescence and 
fragmentation respectively. We notice that at LHC energies the fragmentation is the dominant 
hadronization mechanism to produce the $D^{0}$ meson. 
This is due to coalescence that at high 
energies is less significant, because the effect of the coalescence depends 
on the slope of the charm quark momentum distribution. In fact for an harder charm quark distribution, 
like at LHC, the gain in momentum reflects in a smaller increase in the slope compared to the 
one at RHIC energies, see also Ref. \cite{Scardina:2017ipo}.

\begin{figure}[t]
\centering
\includegraphics[width=\columnwidth, angle=0,clip]{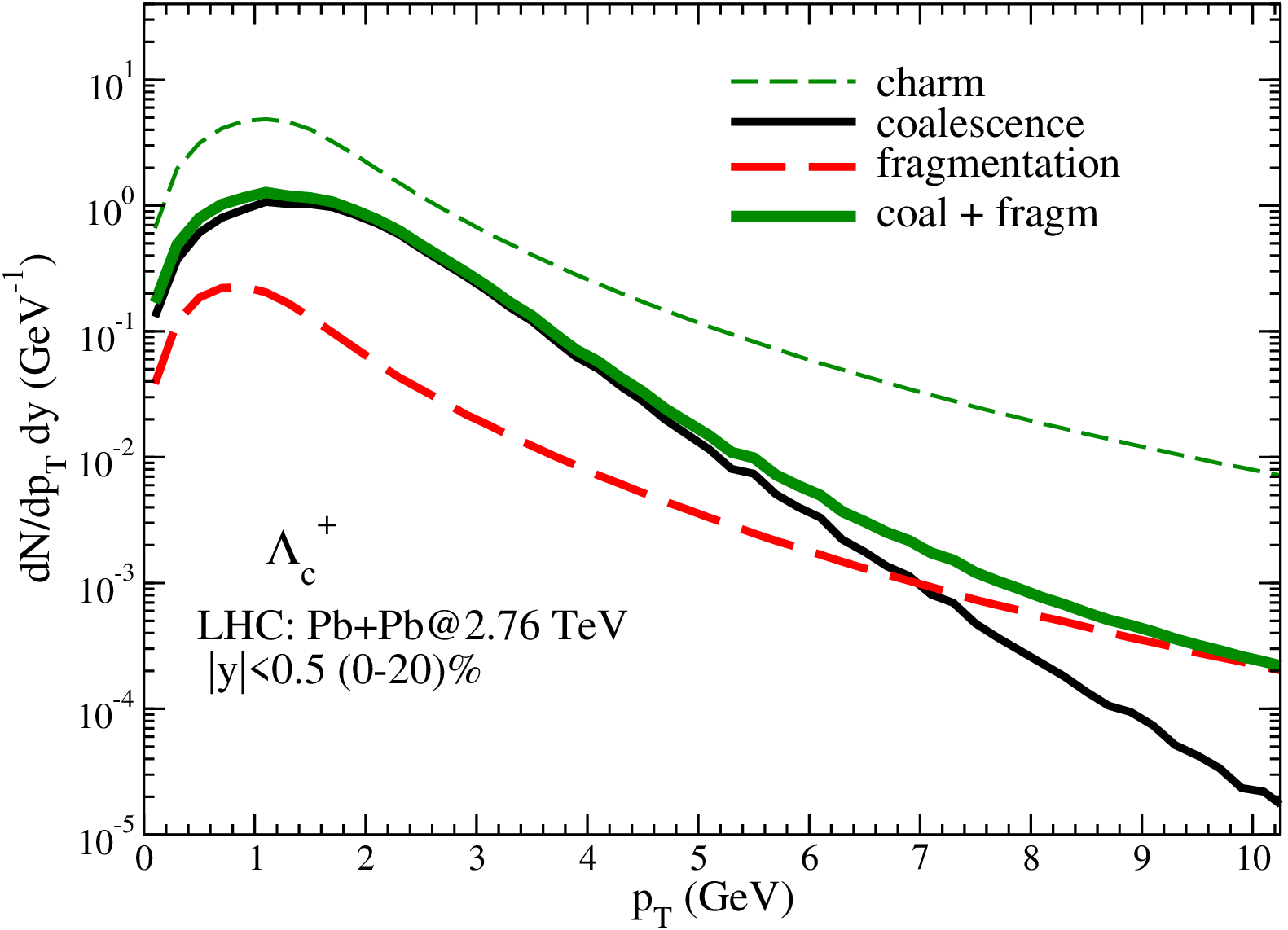}
\caption{\label{Fig:LambdaC_LHC}
(Color online) 
Transverse momentum spectra for $\Lambda_{c}^{+}$ baryon at mid-rapidity 
for $Pb+Pb$ collisions at $\sqrt{s}=2.76 \, \mbox{TeV}$ and for $(0-20\%)$ centrality.
Green dashed line refers to the charm spectrum. Black solid and red dashed lines refer to the $\Lambda_c^{+}$
spectrum from only coalescence and only fragmentation respectively while green solid line refers 
to the sum of fragmentation and coalescence processes. 
}
\end{figure}
The $\Lambda_{c}^{+}$ momentum spectrum at mid-rapidity 
for $(0-20\%)$ centrality is shown in Fig. \ref{Fig:LambdaC_LHC}. 
Also at LHC energies coalescence has the dominant
role for charmed baryon production in the region where $p_{T} <5 \,\rm \mbox{GeV}$. 
The ratio of $\Lambda_{c}$ from coalescence and fragmentation at LHC is smaller than at RHIC, but it remains significant in the region at low momenta.

\begin{figure}[t]
\centering
\includegraphics[width=\columnwidth, angle=0,clip]{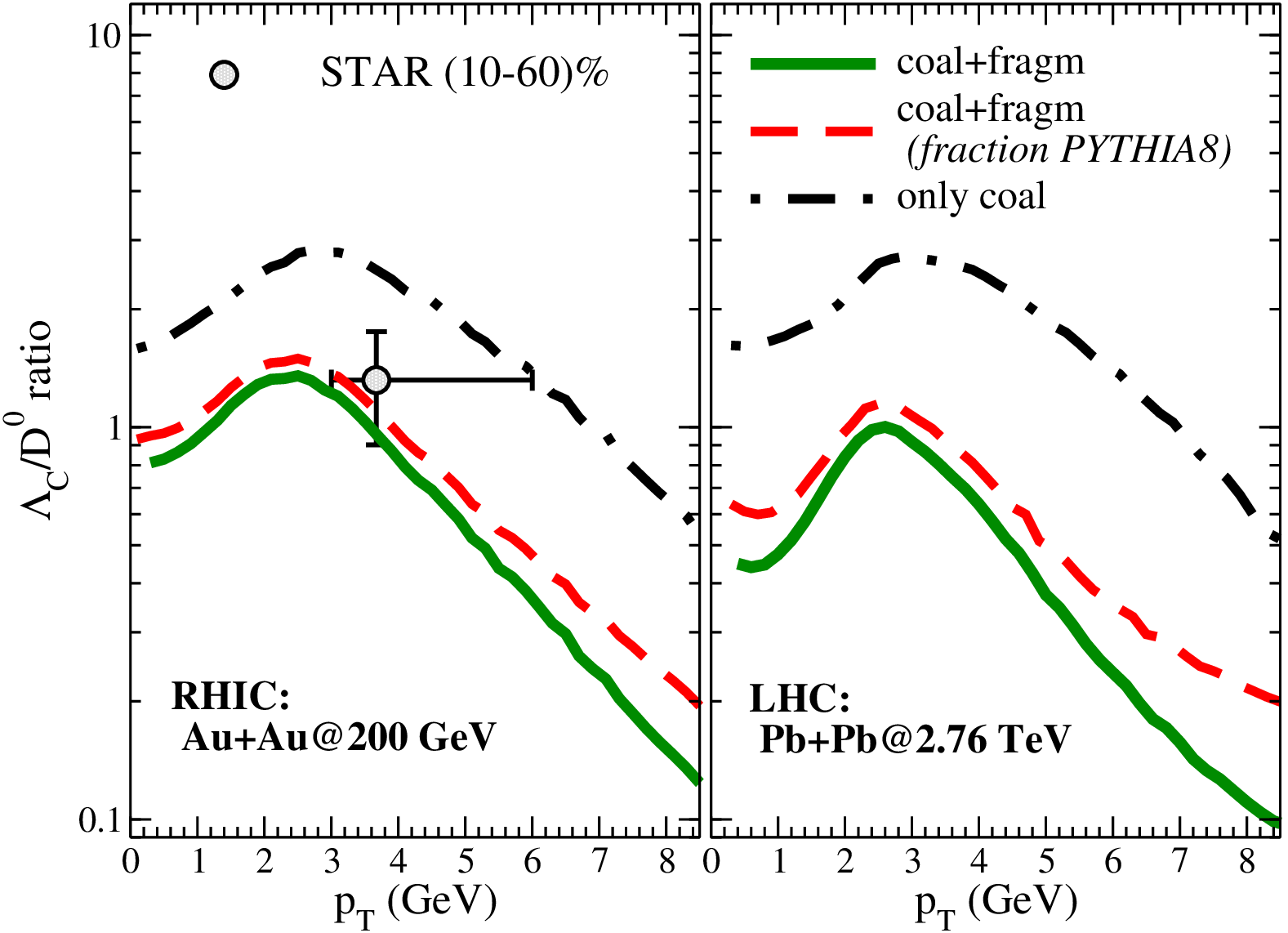}
\caption{\label{Fig:LambdaD0_ratio_LHC} 
(Color online) 
$\Lambda_{c}^{+}$ to $D^0$ ratio as a function of $p_T$ and at mid-rapidity for $Au+Au$ collisions at $\sqrt{s}=200 \, \mbox{GeV}$ (left panel)
and for $Pb+Pb$ collisions at $\sqrt{s}=2.76 \, \mbox{TeV}$ (right panel). Black dot-dashed lines refer to calculations with only coalescence 
while green solid lines refer to coalescence plus fragmentation. The red dashed lines refer to coalescence plus fragmentation but with fragmentation normalized to the fraction of PYTHIA8 \cite{Sjostrand:2014zea}.
Experimental data taken from \cite{Zhou:2017ikn}. 
}
\end{figure}
The comparison of $\Lambda_{c}^{+}/D^0$ ratio as a function of $p_T$ 
between RHIC energies (left panel) and LHC energies (right panel) is shown in Fig. \ref{Fig:LambdaD0_ratio_LHC}.
As we can see comparing dot-dashed lines at both RHIC and LHC energies the coalescence predict similar baryon/meson ratio for both energies.
As described in section 4 the baryon to meson ratio from fragmentation is established from the experimental measured fragmentation fraction into final hadrons channels, and it remains the same changing the collision energy. 
Moreover, we observe that at LHC energies coalescence plus fragmentation predict a smaller $\Lambda_{c}^{+}/D^0$. 
Even if the only coalescence ratio and the only fragmentation ratio remain similar at RHIC and LHC, the combined ratio is different because
the coalescence over fragmentation ratio at LHC is smaller than at RHIC.
Therefore at LHC the larger contribution in particle production from fragmentation leads to a final ratio that is smaller than at RHIC, in fact the ratio from fragmentation fraction is $\sim 0.09$ and the one from coalescence is about 1.

\section{Conclusions}\label{section:Conclusion}

In this paper we have studied the transverse momentum spectra of charmed hadrons 
($D$ mesons and $\Lambda_C$ baryons) and the $\Lambda_C/D^0$ ratio in heavy ion collisions
for RHIC and LHC energies. In particular we have discussed the enhancement of these 
ratios within a covariant coalescence model of heavy quarks with light quarks. 
For the $p_T$ distributions of light partons in the quark-gluon plasma we have used 
a thermal distribution with a temperature similar to the phase transition
temperature, $T_c \simeq 160 \,\rm \mbox{MeV}$ and included the effect of a radial flow $\beta$.
The volume and radial flow of the hadronizing fireball have been constrained 
by the total multiplicity and total transverse energy as done in \cite{Minissale:2015zwa}. 
The core of the approach is the one developed for RHIC energies and recently extended to study also 
LHC energies \cite{Greco:2003mm,Minissale:2015zwa}.
The width parameters of hadron Wigner functions
used in the coalescence model have been determined according to the 
charge radius calculated in the quark models and normalized to have all 
charmed hadrons low $p_T \simeq 0$ formed by coalescence. 
The remaining charm quarks have been converted to heavy hadrons by mean of 
fragmentation as in p+p collisions.
This ensures that in any momentum all the charm quark undergo hadronization.
We have also included the contribution from main hadronic channels including the ground states and the 
first excited states for $D$ and $\Lambda_c$ hadrons in estimating the ratios.

We have studied the $p_T$ spectra evolution from RHIC to LHC energies for the charmed 
hadrons $D$ and $\Lambda_c$. The results obtained are in good agreement with recent 
 $D^0$ mesons experimental data from RHIC and LHC in central collisions. 
Finally, we have studied the  $\Lambda_c/D^0$ ratio $p_T$ dependence at different energies.
The comparison with the light baryon/meson ratios shows that the heavy baryon/meson ratio 
has a weaker dependence on the transverse momentum due to the massive charms quarks inside
heavy hadrons.
We have found that our approach predict $\Lambda_c/D^0 \simeq 1.5$ and it peaks at $p_T \simeq 3 \, \rm \mbox{GeV}$ at 
RHIC energies. 
However it has to be noted that the value of the peak is only about a factor of 2 larger than the value at $p_T \to 0$.
This remains true even if we adjust the $\Lambda_c^{+}$ radius to have a ratio at low $p_T$ 
to be about 0.2 like in thermal models. The underlying reason is that within a coalescence 
mechanism the gain in $p_T$ due to a coalescence with a light quark is modest, so one does not have
the large enhancement from low $p_T$ to intermediate $p_T$ like in the $p/\pi$ and $\Lambda/K$ ratio
\cite{Fries:2003vb,Greco:2003xt,Greco:2003mm,Fries:2003kq,Minissale:2015zwa}

Furthermore, we have found a strong enhancement of heavy baryon over heavy meson 
ratio due to coalescence at low $p_T$ compared to the one from thermal model. In fact, 
coalescence model predicts a $\Lambda_c/D^0 \simeq 0.75$ at RHIC energies
in the $p_{T} \simeq 0$ region where simple thermal model predicts, in the same region, a factor 
$2-3$ smaller for this ratio.
Therefore the $\Lambda_{c}/D^0$ ratio is a good tool to disentangle different hadronization mechanisms 
once the data will be available, mostly in the low $p_T$ regime. 
Finally, we observe that at LHC energies even if coalescence probability is nearly the same 
as at RHIC, 
the relative production w.r.t fragmentation decrease 
leading to predict a slight decrease of the 
$\Lambda_c/D^0$ ratio.


\appendix

\section{Approximate evaluation of coalescence integral}\label{App:coal}
To get an approximate evaluation of the coalescence integral Eq. \ref{eq-coal}
we follow Ref.s \cite{Chen:2007zp,Cho:2011ew,Sun:2017ooe}.
We consider an hypersurface of constant proper time.
Moreover we assume that the particles are uniformly distributed
in space and have momentum distributions given by Boltzmann distribution 
with Bjorken correlation of equal spatial $\eta_i$ and momentum $y_i$ rapidities as follows:
\begin{eqnarray}
f_{i}(x_i,p_i)=g_i \, e^{-\frac{p^{\mu}u_{\mu}}{T}}\delta(y_i-\eta_i) \,\, i=1,..,n\nonumber
\end{eqnarray}
In the following discussion $n$ is the number of the constituent quarks of the hadron.
Therefore the integral is given by
\begin{eqnarray}
\frac{d^{2}N_{H}}{dP_{T}^{2}}&=& g_{H} \int \prod^{n}_{i=1} \tau m_{T i} d^2x_i dy_i d^2p_i \frac{g_i}{(2\pi)^3} e^{-m_{Ti}/T} \times \nonumber \\ 
&\times& f_{H}(x_{1}...x_{n}, p_{1}...p_{n})\, \delta^{(2)} \left(P_{T}-\sum^{n}_{i=1} p_{T,i} \right) \nonumber
\end{eqnarray}
where for this approximate calculation we have neglected the transverse flow of produced matter,
being more interested to point out the parameter dependence of the yield on the masses
of the coalescing quarks.
We introduce the center-of-mass position vector $X_{cm}$ and the relative 
spatial coordinate vectors $x_{r i}$ that they can be expressed as
\begin{eqnarray}
X_{cm} = \frac{\sum_{j=1}^{n}m_{j} x_{j}}{\sum_{j=1}^{n}m_{j}} , \, \, \, x_{r i} = \Bigg(\frac{\sum_{j=1}^{i}m_{j} x_{j}}{\sum_{j=1}^{i}m_{j}}- x_{i+1}\Bigg) \nonumber
\end{eqnarray}
Correspondingly, in the momentum space, we introduce the total momentum $P_{tot}$ 
the relative momentum vectors $p_{r i}$.
With this change of variable we have:
\begin{eqnarray}
\prod_{i=1}^{n} d^2x_i d^2p_i=d^2X_{cm} d^2P_{tot} \prod_{i=1}^{n-1} d^2x_{r i} d^2p_{r i}
\end{eqnarray}
The Wigner function does not depend on the center-of-mass coordinate and it depends 
only on the relative coordinates. 
\begin{eqnarray}
f_H(x_i,p_i)=A^{n-1} \exp{\bigg\{ -\Big( \sum_{i=1}^{n-1} \frac{x_{ri}^{2}}{\sigma_{r i}^2}+\sum_{i=1}^{n-1} p_{ri}^{2} \sigma_{r i}^2\Big)\bigg \}} \nonumber
\end{eqnarray}
where $A$ is the normalization factor.
For the quark distribution function we use the non-relativistic approximation 
and using the relative momentum vectors we have
\begin{eqnarray}
\prod_{i=1}^{n} \exp{\Big[-\frac{m_{Ti}}{T} \Big]} &\simeq& \exp{\Big[ -\frac{M}{T} \Big]} 
\exp{\Big[-\frac{P_{tot}^2}{2 M T}\Big]} \times \nonumber \\
&\times & \exp{\Big[-\sum_{i=1}^{n-1}\frac{p_{ri}^2}{2 \mu_i T}\Big]}
\end{eqnarray}
where $M=\sum_{i=1}^{n} m_i$ is the total mass while $\mu_{i}$ are the reduced masses defined by 
\begin{eqnarray} 
\mu_{i}=\frac{m_{i+1}\sum_{j=1}^{i}m_j}{\sum_{j=1}^{i+1}m_j}  \, \, \, i=1,...,n-1
\label{Eq.App.mu2}
\end{eqnarray}
Notice that the reduced mass have the following property $\prod_{i=1}^{n} m_i=\Big( \sum_{i=1}^{n} m_i \Big) \prod_{i=1}^{n-1}\mu_i$.
In the non relativistic limit
\begin{eqnarray}
\prod_{i=1}^{n} m_{Ti} &\simeq& \Big(\prod_{i=1}^{n}m_i \Big) \Big[ 1+ \sum_{i=1}^{n} \frac{p_i^2}{2 m_i^2}\Big]
\end{eqnarray}
The integrations in the center of mass coordinate and in the total momentum are straightforward and give 
$A_{T} \exp{\big[-P_{T}^2/(2 M T)\big]}$ where $A_{T}$ is the transverse area of the fireball. The integration in the relative 
coordinate are gaussian integration. 
Finally, we obtain the following approximate coalescence formula for the momentum spectra of the hadron at mid rapidity
\begin{eqnarray}
\frac{d^{2}N_{H}}{dP_{T}^{2}}&\simeq&g_{H} A_{T} M e^{-M/T} e^{-P_{T}^2/(2 M T)} A^{n-1} \times \nonumber \\ 
&\times& \Big[ \prod_{i=1}^{n} \frac{\tau g_i}{(2\pi)^3}\Big]  \Big[ \prod_{i=1}^{n-1} \mu_i \Big] \Big[ \prod_{i=1}^{n-1} \pi^2 \sigma_{r i}^2 \xi_i^2\Big]
\end{eqnarray}
where $\xi_{i}=[\sigma_{r i}^2 + 1/(2 \mu_i T)]^{-1}$. Therefore from this formula one can get the baryon-to-meson ratio 
for a case of $[q \, q \, q']/[q' \, \overline{q}]$ for low transverse momentum and assuming for the widths of mesons and baryons 
$\sigma_r=\sigma_{r 1}$ as follows
\begin{eqnarray}
\label{APP:ratio}
\frac{N_{B}}{N_{M}}\bigg|_{P_T\simeq0} &\simeq& \frac{g_{B}}{g_{M}} \Big( \frac{M_B}{M_M}\Big) e^{-(M_B-M_M)/T} \, 
A_W \rho_q \sigma_{r 2}^2 \, \xi_2^2 \, \frac{\mu_2}{m_q} \nonumber \\
\end{eqnarray}
where $\rho_q$ is the quark density in the transverse plane.
Therefore, the Baryon-to-meson ratio shows a first term similar to the one of the thermal model proportional to 
$\Big( \frac{M_B}{M_M}\Big) e^{-(M_B-M_M)/T}$ and a second term that take into account for microscopic 
details of the hadronization mechanism that depends on the reduced mass $\mu_2$.
This means that the baryon-to-meson ratio increase with the increasing of the reduced mass of the baryon $\mu_2$. 
In other words at low $p_T$ coalescence predict a mass ordering for the baryon-to-meson ratio.


\end{document}